\documentclass[12pt,preprint]{aastex}

\shorttitle{Asymmetric SNe and Highly Collimated GRBs}
\shortauthors{Tsui and Navia}

\begin{document}

\title{Plasma Pressure Driven Asymmetric Supernovae
\\
 and Highly Collimated Gamma-Ray Bursts}
\author{K.H. Tsui and C.E. Navia}
\affil{Instituto de F\'{i}sica - Universidade Federal Fluminense
\\Campus da Praia Vermelha, Av. General Milton Tavares de Souza s/n
\\Gragoat\'{a}, 24.210-346, Niter\'{o}i, Rio de Janeiro, Brasil.}
\email{tsui$@$if.uff.br}
\pagestyle{myheadings}
\baselineskip 18pt
	 
\begin{abstract}

During the process of collapse of a massive star,
 a cavity is generated between the central iron core
 and an outer stellar envelope.
 The dynamics of this cavity,
 filled with plasma and magnetic field
 of the rapidly rotating proto-magnetar's magnetosphere,
 is believed to be very relevant
 in understanding supernovae and gamma-ray bursts.
 The interactions of the pressurized
 conducting plasma and the magnetic fields
 are described by a set of magnetohydrodynamic (MHD) equations
 with poloidal and toroidal plasma flows
 not aligned with magnetic fields.
 A sequence of MHD equilibria
 in response to the increasing plasma pressure in the cavity,
 by continuous filling from the rotating magnetosphere,
 is solved to account for asymmetric supernovae,
 highly collimated gamma-ray burst jets,
 and also active galactic nucleus plasma torus.
 It is shown that the magnetosphere
 of the central compact star
 is likely the central engine
 of supernova and gamma-ray burst
 by feeding them plasma, magnetic energy,
 and rotational energy.

\end{abstract}

\maketitle

\newpage
\section{Introduction}

Ever since Baade and Zwicky had coined the term super-novae
 and had suggested super-nova process
 as the end of an ordinary star
 to become a neutron star \citep{baade1934},
 supernovae have been puzzling scientific minds for decades.
 They are believed to be the catastrophic end
 of massive stars' life cycle.
 Supernovae other than the type Ia standard candle,
 where the entire star is incinerated,
 are believed to have gone through a gravitational collapse
 of the iron core \citep{bethe1990, kotake2006}.
 The progenitor iron core with about $10^{4}\,Km$ radius
 is enclosed by the stellar hydrogen envelope
 extending out to some $10^{8}\,Km$ radius.
 The collapse of this progenitor core
 is governed by the sound speed profile,
 which decreases as the radial position increases.
 The infalling velocity in the outer part of the core
 is supersonic with respect to the local sound speed,
 while the inner part is subsonic
 with the interface at about $300\,Km$ radius.
 Because of this condition,
 the outer part tends to pile up
 while the inner part continues to free-fall.
 This generates a proto-neutron star
 plus the overlying stellar material.
 A cavity is formed between the proto-neutron star
 and the hydrogen stellar envelope.
 The gravitational collapse of a magnetized massive star
 with jet formations was investigated \citep{leblanc1970},
 and an explosion mechanism not associated
 with nuclear detonation was proposed
 \citep{bisnovatyi1971}.

Some of the core-collapse supernovae
 appear to be associated to gamma-ray bursts
 \citep{galama1999, campana2006, pian2006, soderberg2006, mazzali2006}.
 As a result, there could be a very close relationship
 between supernovae and gamma-ray bursts.
 By the extraordinary energy output,
 serveral scenarios have been contemplated
 as the progenitors of gamma-ray bursts \citep{meszaros2002}.
 For massive progenitor stars with more than 14 solar masses,
 like rapidly spinning Wolf-Rayet stars,
 the inner core could promptly collapse to a black hole
 circumscribed by a massive torus,
 failing to generate a core rebound in the cavity.
 Accretion of the surrounding massive torus
 of nuclear density material at a later time
 drive an outburst along the rotational axis
 breaking out the stellar envelope.
 This is the Collapsar (or Hypernova) scheme \citep{macfadyen1999}.
 Instead of promptly collapsing to a black hole,
 there also could be a two-stage collapse
 by first forming a rapidly rotating neutron star
 temporarily stabilized by rotation,
 and later collapses to a black hole
 after losing some angular momentum,
 which is the Supranova \citep{vietri1999} scheme
 for long gamma-ray bursts.
 When the inner core does not collapse to a black hole,
 it rebounces as the inner core reaches nuclear densities,
 generating a rebounding shock in the cavity.
 The release of the gravitational binding energy
 through neutrino bursts has been considered
 as the primary candidate in fueling the supernova explosion
 \citep{bethe1990, kotake2006, bethe1985}.
 Nevertheless, as this outgoing shock
 meets the infalling outer very thick envelope,
 energy of the shock wave gets dissipated,
 and the shock is stalled.
 By which energy source and mechanism
 that the shock could be reignited is still an unsettled issue.
 
Recently, polarimetry observations of supernova optical emissions
 have revealled different degrees of polarization
 along a fixed axis of the supernova.
 In a very collisional environment of a supernova,
 this polarization result implies a nonzero volume average
 of the microscopically random electric field vector of each emission,
 indicating supernovae, or some them, are aspherical
 \citep{howell2001, wang2003}.
 In view of other astrophysical ejection events
 known to be powered by magnetic fields,
 these polarization observations
 have given grounds to reexamine the magnetic field
 as the central engine for supernova
 \citep{wheeler2000, wheeler2002, ardeljan2005,
 uzdensky2007, burrows2007, komissarov2007}.
 The primary concern in this renewed magnetic approach
 is the magnetic collimation mechanism.
 It has been proposed the presence of an accretion disk
 within the cavity of the collapsing iron core,
 by which magnetic towering \citep{lynden2003}
 and jets \citep{blandford1982}
 could be generated along the rotational axis.

Should magnetic field be the energy source of supernova,
 a simple estimation indicates
 that the surface fields of the rapidly rotating
 proto-neutron star would be around $10^{15}\,Gs$,
 which qualifies it as a magnetar.
 Such magnetar scenario could be acomplished
 if the rotation of the proto-neutron star is fast enough.
 With millisecond periods,
 dynamo effects inside the pulsar could be triggered,
 enhancing the magnetic fields by a factor of $10^{3}$
 \citep{duncan1992, thompson1993}.
 These magnetic fields would be launched
 from the magnetosphere together with the plasma
 to fill the cavity,
 leaving a normal spun-down neutron star at the center
 after supernova explosion.
 In general, the dynamics of this cavity
 driven by the rapidly rotating inner core
 is referred to as the 'Pulsar in a Cavity' model
 \citep{uzdensky2007}.
 Simulation has been the principle tool
 in investigating the dynamics in this cavity
 \citep{burrows2007, komissarov2007}.
 Most frequently, an accretion disk scenario
 with magnetic fields anchored on it is considered.
 Under given initial conditions,
 magnetic towering \citep{lynden2003}
 due to the angular momentum of the disk
 generates a collimating magnetic column
 along the polar axis.

Here, we seek to describe the cavity structures,
 filled with plasma and magnetic fields,
 through a sequence of axisymmetric
 magnetohydrodynamic (MHD) steady states (equilibria)
 with poloidal and toroidal plasma rotations
 as a response to the increasing plasma pressure.
 This type of steady state analysis
 has been used to gain important insights
 of dynamic processes
 such as astrophysical jets in terms of
 spatially self-similar MHD equilibria
 \citep{blandford1982}
 and magnetic towering in terms of
 magnetostatic analysis of disk driven equilibria
 \citep{lynden2003}.
 The present problem differs from
 the relativistic pulsar wind problem
 with hot plasma \citep{lovelace1986, camenzind1986}
 especially under self-similar analysis
 \citep{prendergast2005, gourgouliatos2010}
 and with cold plasma
 \citep{okamoto1978, begelman1994, okamoto2002}
 in that the spatial domain is bounded and very finite.
 
In Sec.2, the axisymmetric divergence-free rotational MHD
 formulation is briefly presented,
 and in Sec.3, by assigning two source functions,
 the Grad-Shafranov equation
 for rotational equilibrium is obtained.
 The nonlinear poloidal flux function is solved in Sec.4
 for asymmetric supernova configuration in the cavity,
 whereas cusp-like funnel polar collimated
 gamma-ray burst jet configuration
 is solved in Sec.5.
 Active galactic nucleus plasma torus
 and some conclusions are finalized in Sec.6.

\newpage
\section{Divergence-Free Rotational MHD}

The standard steady state MHD equations are

\begin{eqnarray}
\nonumber
\nabla\cdot(\rho\vec v)\,&=&\,0\,\,\,,
\\
\nonumber
\rho(\vec v\cdot\nabla)\vec v\,&
 =&\,\vec J\times\vec B-\nabla p-\rho{GM\over r^3}\vec r\,\,\,,
\\
\nonumber
\vec v\times\vec B\,&=&\,-\vec E\,\,\,,
\\
\nonumber
\nabla\times\vec B\,&=&\,\mu\vec J\,\,\,,
\\
\nonumber
\nabla\cdot\vec B\,&=&\,0\,\,\,,
\\
\nonumber
(\vec v\cdot \nabla)\left({p\over\rho^{\gamma}}\right)\,&=&\,0\,\,\,.
\end{eqnarray}

\noindent Here, $\rho$ is the mass density, $\vec v$ is the bulk velocity,
 $\vec J$ is the current density, $\vec B$ is the magnetic field,
 $p$ is the plasma pressure, $M$ is the mass of the central body,
 $\mu$ is the free space permeability,
 and $\gamma$ is the polytropic index.
 To describe plasma equilibrium with poloidal and toroidal flows,
 this set of equations is quite inconvenient.
 We will use the MHD equations for divergence-free plasma flows,
 where the plasma velocity is density weighed
 so that the plasma density appears
 only through the density weighed velocity
 $\vec w_{*}=(\mu\rho)^{1/2}\vec v$,
 except in the gravity term \citep{tsui2011, tsui2012}.

\begin{eqnarray}
\label{eqno1}
(\mu\rho)^{1/2}\nabla\cdot\vec w_{*}
 +\vec w_{*}\cdot\nabla(\mu\rho)^{1/2}\,&
 =&\,0\,\,\,,
\\
\label{eqno2}
\vec w_{*}\times\nabla\times\vec w_{*}
 -\vec B\times\nabla\times\vec B\,&
 =&\,\nabla \mu p_{*}-\rho{GM\over r^3}\vec r\,\,\,,
\\
\label{eqno3}
\vec w_{*}\times\vec B\,&
 =&\,(\mu\rho)^{1/2}\nabla\tilde\Phi\,
 =\,\nabla\Phi\,\,\,,
\\
\label{eqno4}
\nabla\times\vec B\,&=&\,\mu\vec J\,\,\,,
\\
\label{eqno5}
\nabla\cdot\vec B\,&=&\,0\,\,\,,
\\
\label{eqno6}
\nabla\cdot\vec v\,&=&\,0\,\,\,,
\end{eqnarray}

\noindent where $\mu p_{*}=\mu p+w_{*}^{2}/2$
 is the total plasma pressure.
 With divergence-free flows of Eq.~\ref{eqno6},
 density flux conservation gives $\vec v\cdot\nabla\rho=0$.
 By using the $(\vec w_{*},\mu p_{*})$ representation,
 this becomes $\vec w_{*}\cdot\nabla(\mu\rho)^{1/2}=0$,
 and Eq.~\ref{eqno1} then becomes $\nabla\cdot\vec w_{*}=0$.
 Furthermore, making use of Eq.~\ref{eqno4},
 the $\vec w_{*}$ term and the $\vec B$ term
 are in symmetric form in Eq.~\ref{eqno2}.
 Because of Eq.~\ref{eqno5},
 the magnetic field can be represented
 through a vector potential.
 Under axisymmetry, this vector potential
 allows the magnetic field
 be represented by two scalar functions,
 which reads

\begin{eqnarray}
\label{eqno7}
\vec B\,
 &=&\,A_{0}(\nabla\Psi\times\nabla\phi+F\nabla\phi)\,
 =\,{A_{0}\over r\sin\theta}
 \left(+{1\over r}{\partial\Psi\over\partial\theta},
   -{\partial\Psi\over\partial r},
   +F\right)\,\,\,,
\\
\nonumber
\mu\vec J\,
 &=&\,{A_{0}\over r\sin\theta}
 \left(+{1\over r}{\partial F\over\partial\theta},
   -{\partial F\over\partial r},
   -{\partial^2\Psi\over\partial r^2}
   -{1\over r^2}\sin\theta
   {\partial\over\partial\theta}
   ({1\over\sin\theta}{\partial\Psi\over\partial\theta})\right)
   \,\,\,.
\end{eqnarray}

\noindent Here, $A_{0}$ carries the physical dimension
 of poloidal magnetic flux
 such that $\Psi$ is a dimensionless poloidal flux function,
 and $A_{0}F$ is a measure of the axial plasma current.
 Likewise, because $\nabla\cdot\vec w_{*}=0$ by Eq.~\ref{eqno1},
 we also have

\begin{eqnarray}
\label{eqno8}
\vec w_{*}\,=\,A'_{0}(\nabla\Psi'\times\nabla\phi+F'\nabla\phi)\,\,\,.
\end{eqnarray}

\noindent We note that $A_{0}$ and $A'_{0}$
 are not the maximum values of poloidal magnetic
 and density weighed velocity fluxes.
 They are reference vlues only
 as such that the dimensionless poloidal flux functions
 are not normalized to unity.
 The condition $\vec w_{*}\cdot\nabla\rho=0$
 gives $\rho=\rho(\Psi')$.
 The scalar product of $\vec w_{*}$ on Eq.~\ref{eqno3}
 results in $\tilde\Phi=\tilde\Phi(\Psi')$.
 Consequently, with $\rho=\rho(\Psi')$,
 the right side of Eq.~\ref{eqno3}
 can be written as the gradient of $\Phi=\Phi(\Psi')$ only
 giving the second equality.
 Taking the scalar product again of $\vec B$ and $\vec w_{*}$
 on Eq.~\ref{eqno3} give $\Phi=\Phi(\Psi)$ and $\Phi=\Phi(\Psi')$
 respectively thus $\Psi'=\Psi'(\Psi)$.
 Substituting Eq.~\ref{eqno7} and Eq.~\ref{eqno8} to Eq.~\ref{eqno3},
 gives
 
\begin{eqnarray}
\nonumber
(\nabla\Psi'\times\nabla\Phi)\times(\nabla\Psi\times\nabla\Phi)
 -F\nabla\Phi\times(\nabla\Psi'\times\nabla\Phi)
 +F'\nabla\Phi\times(\nabla\Psi\times\nabla\Phi)\,
\\
\nonumber
 =\,{1\over A_{0}}{1\over A'_{0}}\nabla\Phi\,\,\,.
\end{eqnarray}
 
\noindent Taking the scalar product of $\nabla\Psi$
 on this equation gives

\begin{mathletters}
\begin{eqnarray}
\label{eqno9a}
{1\over A_{0}}{1\over A'_{0}}\nabla\Psi\cdot\nabla\Phi\,
 &=&\,(\nabla\phi)^{2}\nabla\Psi\cdot(F'\nabla\Psi-F\nabla\Psi')
 \,\,\,,
\\
\label{eqno9b}
\left(F'-F{\partial\Psi'(\Psi)\over\partial\Psi}\right)\,
 &=&\,{1\over A_{0}A'_{0}}{1\over(\nabla\phi)^{2}}
 {\partial\Phi(\Psi)\over\partial\Psi}\,\,\,.
\end{eqnarray}
\end{mathletters}

\newpage
\section{Rotational Grad-Shafranov Equation}

As for Eq.~\ref{eqno2}, with $\Psi'=\Psi'(\Psi)$,
 the $\phi$ component gives

\begin{eqnarray}
\nonumber
A'^{2}_{0}{\partial\Psi'\over\partial\Psi}
 \left(-{\partial\Psi\over\partial r}{\partial F'\over\partial\theta}
 +{\partial\Psi\over\partial\theta}{\partial F'\over\partial r}\right)
 -A_{0}^{2}
 \left(-{\partial\Psi\over\partial r}{\partial F\over\partial\theta}
 +{\partial\Psi\over\partial\theta}{\partial F\over\partial r}\right)\,
 =\,0\,\,\,.
\end{eqnarray}

\noindent We use Eq.~\ref{eqno9b} to eliminate $F'$.
 It is noted that $F$ and $F'$ are not functions of $\Psi$
 only for rotational equilibrium
 due to the $(\nabla\phi)^{2}$ factor.
 With $(\nabla\phi)^{2}=(1/r\sin\theta)^{2}$,
 we then get

\begin{eqnarray}
\nonumber
\left[A'^{2}_{0}({\partial\Psi'\over\partial\Psi})^{2}-A_{0}^{2}\right]
 \left(-{\partial\Psi\over\partial r}{\partial F\over\partial\theta}
 +{\partial\Psi\over\partial\theta}{\partial F\over\partial r}\right)
\\
\nonumber
 +{A'_{0}\over A_{0}}{\partial\Psi'\over\partial\Psi}
 {\partial\Phi\over\partial\Psi}
 \left(-{\partial\Psi\over\partial r}
 {\partial\over\partial\theta}(r^{2}\sin^{2}\theta)
 +{\partial\Psi\over\partial\theta}
 {\partial\over\partial r}(r^{2}\sin^{2}\theta)\right)\,
 =\,0\,\,\,.
\end{eqnarray}

\noindent By inspection, we have

\begin{mathletters}
\begin{eqnarray}
\label{eqno10a}
F(r,\theta)\,
 =\,k(ar)^{2}\sin^{2}\theta\,\,\,,
\\
\label{eqno10b}
\left[({\partial\Psi'\over\partial\Psi})^{2}
 -({A_{0}\over A'_{0}})^{2}\right]
 +{1\over ka^{2}}{1\over A'_{0}}{1\over A_{0}}
 {\partial\Psi'\over\partial\Psi}
 {\partial\Phi\over\partial\Psi}\,
 =\,0\,\,\,.
\end{eqnarray}
\end{mathletters}

\noindent We note that $a$ is a normalizing factor of $r$,
 and $k$ is an inverse scale length
 such that the scalar function $F(r,\theta)$
 has the correct dimension of $1/r$ in Eq.~\ref{eqno7}.
 An arbitrary constant could be added to Eq.~\ref{eqno10a}.
 However, this constant would correspond to
 an externally applied current along the axial direction,
 which would be appropriate for laboratory tokamak plasmas
 but not for astrophysical plasmas.
 Substituting $\partial\Phi/\partial\Psi$ of Eq.~\ref{eqno10b}
 to Eq.~\ref{eqno9b} gives

\begin{eqnarray}
\label{eqno11}
F'(r,\theta)\,
 =\,{\partial\Psi'\over\partial\Psi}F(r,\theta)
 +{1\over ka^{2}}{1\over A'_{0}}{1\over A_{0}}
 {\partial\Phi\over\partial\Psi}F(r,\theta)
 =\,{1\over(\partial\Psi'/\partial\Psi)}
 \left({A_{0}\over A'_{0}}\right)^{2}F(r,\theta)\,\,\,.
\end{eqnarray}

\noindent Choosing the following linear source function leads to

\begin{mathletters}
\begin{eqnarray}
\label{eqno12a}
\Psi'(\Psi)\,&=&\,b\Psi\,\,\,,
\\
\label{eqno12b}
F'(F)\,&=&\,{1\over b}
 \left({A_{0}\over A'_{0}}\right)^{2}F\,
 =\,\alpha F\,\,\,.
\end{eqnarray}
\end{mathletters}

\noindent We note that $b$ is an independent model parameter,
 but $\alpha$, defined in Eq.~\ref{eqno12b},
 is determined by $b$ together with
 the parameters $(A_{0}, A'_{0})$.

In order to analyse the $\theta$ and $r$ components,
 we choose to write the generalized total pressure
 $\bar p_{*}$ in separable form
 
\begin{eqnarray}
\label{eqno13}
\mu\bar p_{*}(r,\theta)\,=\,\mu\bar p_{*}(r,\Psi)\,
 =\,\mu\bar p_{0}\bar p_{*1}(r)\bar p_{*2}(\Psi)\,\,\,,
\end{eqnarray}

\noindent where $\bar p_{0}$ has the physical dimension
 of pressure $\bar p_{*}$,
 and $\bar p_{*1}$ and $\bar p_{*2}$
 are dimensionless functions of $r$ and $\Psi$
 not normalized to unity,
 and $\bar p_{*}$ is defined by 

\begin{equation}
\label{eqno14}
{\mu\bar p_{*}\over ka^{2}}\,
 =\,\left[{\mu p_{*}\over ka^{2}}
 +(A^{2}_{0}-A'^{2}_{0}\alpha^{2})F\right]\,
 =\,\left[{1\over ka^{2}}(\mu p+{1\over 2}w_{*}^{2})
 +(A^{2}_{0}-A'^{2}_{0}\alpha^{2})F\right]\,\,\,.
\end{equation}

\noindent The $\theta$ and $r$ components then become
 
\begin{eqnarray}
\label{eqno15}
\left[r^{2}{\partial^2\Psi\over\partial r^2}
 +\sin\theta{\partial\over\partial\theta}
 \left({1\over\sin\theta}{\partial\Psi\over\partial\theta}\right)\right]\,
 =\,{r^{2}\over ((bA'_{0})^{2}-A^{2}_{0})}{F\over ka^{2}}
 {\partial\over\partial\Psi}(\mu\bar p_{*})\,\,\,,
\end{eqnarray}

\begin{eqnarray}
\nonumber
{\partial\Psi\over\partial r}
 \left[r^{2}{\partial^2\Psi\over\partial r^2}
 +\sin\theta{\partial\over\partial\theta}
 \left({1\over\sin\theta}{\partial\Psi\over\partial\theta}\right)\right]\,
\\
\nonumber
 =\,{r^{2}\over ((bA'_{0})^{2}-A^{2}_{0})}{F\over ka^{2}}
 {\partial\over\partial r}(\mu\bar p_{*})
 +{r^{2}\over ((bA'_{0})^{2}-A^{2}_{0})}{F\over ka^{2}}
 {\partial\Psi\over\partial r}
 {\partial\over\partial\Psi}(\mu\bar p_{*})
\\
\label{eqno16}
 +{r^{2}\over ((bA'_{0})^{2}-A^{2}_{0})}
 r^{2}\sin^{2}\theta\mu\rho{GM\over r^{2}}\,\,\,.
\end{eqnarray} 

\noindent The right side of the $r$ component correspond to
 the explicit derivative $\partial/\partial r$,
 the implicit derivative $(\partial\Psi/\partial r)(\partial/\partial\Psi)$
 of generalized total plasma pressure,
 and the gravitational term.
 Making use of the $\theta$ component
 to eliminate the implicit derivative term,
 the $r$ component becomes
 
\begin{eqnarray}
\label{eqno17}
\mu\bar p_{0}a{\partial\bar p_{*1}\over\partial z}\bar p_{*2}(\Psi)
 +\mu\rho(\Psi){GMa^{2}\over z^{2}}\,=\,0\,\,\,,
\end{eqnarray}

\noindent where we have defined a normalized radial coordinate $z=ar$.
 To satisfy this equation with $\rho=\rho(\Psi)$,
 we take $\bar p_{*1}(z)$ as below to get

\begin{eqnarray}
\label{eqno18}
\bar p_{*1}(z)\,&=&\,{1\over z}\,\,\,,
\\
\label{eqno19}
\rho\,&
 =&\,{\bar p_{0}\over GMa}\bar p_{*2}(\Psi)\,
 =\,\rho_{0}\bar p_{*2}(\Psi)\,
 =\,\rho_{0}\bar\rho_{*2}(\Psi)\,\,\,.
\end{eqnarray}

\noindent This shows that $\rho$ is a funtion of $\Psi$ only,
 consistent to the divergence-free flows,
 with $\rho_{0}$ as the amplitude
 and $\bar\rho_{*2}(\Psi)$ as the dimensionless functional dependence.
 
To solve the $\theta$ component,
 let us further specify the second source function
 $\bar p_{*2}$ as

\begin{eqnarray}
\label{eqno20}
\bar p_{*2}(\Psi)\,=\,(C\pm |\Psi|^{2m})\,>\,0\,\,\,.
\end{eqnarray}

\noindent Since plasma pressure is positive definite,
 $\Psi^{2m}$ is tken in absolute value for any $m$.
 The $\theta$ component then reads

\begin{eqnarray}
\label{eqno21}
r^{2}{\partial^2\Psi\over\partial r^2}
 +\sin\theta{\partial\over\partial\theta}
 \left({1\over\sin\theta}{\partial\Psi\over\partial\theta}\right)
 \pm m\beta_{p}z^{4}\sin^{2}\theta\bar p_{*1}(z)|\Psi|^{2m-1}\,
 =\,0\,\,\,,
\end{eqnarray}

\begin{eqnarray}
\nonumber
\beta_{p}\,
 =\,{2\mu\bar p_{0}\over (A^{2}_{0}-(bA'_{0})^{2})a^{4}}\,\,\,.
\end{eqnarray}

\noindent This is the rotational Grad-Shafranov equation
 for divergence-free rotational flows
 with $\beta_{p}$ as the poloidal plasma $\beta$.
 We remark that the static case
 can be recovered by taking $A'_{0}=0$
 and with the corresponding limit of
 $\mu\bar p_{*}$ and $\beta_{p}$.
 This equation in spherical coordinates
 describes the structures in the cavity
 resulting from the interactions
 between the conducting plasma and magnetic fields.
 The corresponding equation without rotational flows
 has been derived by many authors \citep{lovelace1986}
 in cylindrical coordinates.

\newpage
\section{Asymmetric Supernovae}

Writing $\Psi(r,\theta)=R(r)\Theta(\theta)$ in separable form,
 the rotational Grad-Shafranov equation becomes
 
\begin{eqnarray}
\label{eqno22}
r^{2}{1\over R}{d^2R\over dr^2}
 +{1\over \Theta}\sin\theta{d\over d\theta}
 \left({1\over\sin\theta}{d\Theta\over d\theta}\right)
 \pm m\beta_{p}z^{4}\sin^{2}\theta
 \bar p_{*1}(z)R^{2m-2}|\Theta|^{2m-2}\,
 =\,0\,\,\,.
\end{eqnarray}

\noindent We will arrange the above equation in the following form
 to solve the poloidal magnetic flux function $\Psi$
 by separation of variables

\begin{eqnarray}
\nonumber
r^{2}{1\over R}{d^2R\over dr^2}\,
 &=&\,-{1\over \Theta}\sin\theta{d\over d\theta}
 \left({1\over\sin\theta}{d\Theta\over d\theta}\right)
 \mp m\beta_{p}z^{4}\sin^{2}\theta
 \bar p_{*1}(z)R^{2m-2}|\Theta|^{2m-2}\,
\\
\label{eqno23}
 &=&\,n(n+1)\,\,\,.
\end{eqnarray}

\noindent The $R$ equation gives $R(z)=1/z^{n}$
 and $R(z)=z^{n+1}$ as two independent solutions.
 As for the $\Theta$ part, in order to be separable,
 we choose $R(z)=1/z^{n}$ and $\bar p_{*1}(z)=1/z$ to get

\begin{eqnarray}
\nonumber
\sin\theta{d\over d\theta}
 \left({1\over\sin\theta}{d\Theta\over d\theta}\right)
 +n(n+1)\Theta\,
 &=&\,\mp m\beta_{p}z^{3}\left({1\over z^{n}}\right)^{2m-2}
 \sin^{2}\theta|\Theta|^{2m-1}\,
\\
\nonumber
 &=&\,\mp m(n)\beta_{p}\sin^{2}\theta|\Theta|^{2m(n)-1}\,\,\,,
\end{eqnarray}

\noindent with $2n(m-1)=3$ or $m(n)=(3+2n)/2n$.
 Defining $x=\cos\theta$, we have

\begin{eqnarray}
\label{eqno24}
(1-x^2){d^2\Theta(x)\over dx^2}
 +n(n+1)\Theta(x)\,
 =\,\mp m(n)\beta_{p}(1-x^{2})|\Theta|^{2m(n)-1}\,\,\,.
\end{eqnarray}

\noindent This nonlinear equation with $\beta_{p}\neq 0$
 usually would have asymmetric solutions of $\Theta(x)$.
 Nevertheless, we note that this equation is symmetric with $\pm x$,
 hence symmetric nonlinear solutions of $\Theta(x)$ are also allowed.
 As for the magnetic field components,
 they are now given by 

\begin{mathletters}
\begin{eqnarray}
\label{eqna25a}
B_{r}\,&=&\,-{A_{0}a^{2}\over z}
 {1\over z}{\partial\Psi\over\partial x}
 \,\,\,,
\\
\label{eqno25b}
B_{\theta}\,&=&\,-{A_{0}a^{2}\over z}
 {1\over (1-x^{2})^{1/2}}{\partial\Psi\over\partial z}
 \,\,\,,
\\
\label{eqno25c}
B_{\phi}\,&=&\,+{A_{0}a^{2}\over z}
 {1\over (1-x^{2})^{1/2}}{1\over a}F
 \,\,\,.
\end{eqnarray}
\end{mathletters}

\noindent These magnetic fields are expressed
 in terms of the reference field $a^{2}A_{0}$.
 We remark that the $B_{\theta}$ and $B_{\phi}$ fields
 are scaled by the $(1-x^2)^{1/2}$ factor in the denominator,
 which is singular at $x=\pm 1$ or $\theta=0,\pi$.
 Since this singularity is quadratically not integrable,
 the magnetic energy would diverge at this location.
 To eliminate this divergence,
 we require the oscillating function
 $\Theta(x)$ be null at the poles $x=\pm 1$.
 As a result, the oscillating $B_{\theta}$ and $B_{\phi}$
 fields grow to large amplitude near the poles,
 before they plunge to zero at the poles.
 The constraint that $\Theta(\pm 1)=0$
 makes $\beta$ the eigenvalue of  Eq.~\ref{eqno24}.
 The magnetic field lines and, in particular,
 the poloidal field lines are described by

\begin{mathletters}
\begin{eqnarray}
\label{eqno26a}
{B_{r}\over dr}\,
 =\,{B_{\theta}\over rd\theta}\,
 =\,{B_{\phi}\over r\sin\theta d\phi}\,\,\,,
\\
\label{eqno26b}
\Psi(z,x)\,
 =\,R(z)\Theta(x)\,
 =\,C\,\,\,,
\end{eqnarray}
\end{mathletters}

\noindent where the poloidal field lines
 are given by the contours
 of $\Psi(z,x)$ on the $(r-\theta)$ plane,
 which is also shared by the poloidal velocity stream lines.

To solve Eq.~\ref{eqno24}, we note that, for $\beta_{p}=0$,
 $\Theta(x)$ is given in terms of Legendre polynomials
 by Eq.~\ref{eqno29b} in Sec.4 with $n$ axisymmetric lobes,
 as shown in Fig.1 with $n=1$ and $n=3$.
 However, $\beta_{p}=0$ would imply $\bar p_{*}=0$,
 or equivalently $p=0$, $w_{*}=0$, and $A_{0}=0$ by Eq.~\ref{eqno14},
 which amounts to no pressure, no flow, and no magnetic fields.
 For $\beta_{p}\neq 0$, Eq.~\ref{eqno24}
 defines the nonlinear $\Theta(x)$ for a given $n$.
 To solve for equilibrium $\Theta(x)$,
 we take the lower sign in Eq.~\ref{eqno20} and Eq.~\ref{eqno24}
 with $\bar p_{*2}(\Psi)=(C-|\Psi|^{2m})$
 to have a balance with the poloidal magnetic pressure.
 We start from $x=-1$ with $\Theta(-1)=0$
 and a slope $\Theta'(-1)$
 as such to reach $x=+1$ with $\Theta(+1)=0$.
 In the parameter space of $(\Theta'(-1),\beta_{p})$ for a given $n$,
 there will be (n-1)-lobe, (n-2)-lobe, ..., 1-lobe
 structures in general.
 Rather than covering the $(\Theta'(-1),\beta_{p})$ parameter space,
 we take the derivative of the linear $\beta_{p}=0$ solution
 of Eq.~\ref{eqno29b} as the reference slopes at $x=-1$, with

\begin{eqnarray}
\label{eqno27}
\Theta'(x,n)\,=\,-2x{dP_{n}\over dx}\,\,\,.
\end{eqnarray}

\noindent For $n=2$, a 1-lobe structure is shown in Fig.2
 with $\Theta'(-1,1)$ and its corresponding $\beta_{p}=9.57$,
 where the lobe asymmetry is not evident.
 For $n=3$, Fig.3 shows a 2-lobe structure
 with $\Theta'(-1,3)$ and $\beta_{p}=6.69$,
 and for $n=3$ in Fig.4 is another 2-lobe structure
 with a different boundary derivative $\Theta'(-1,2)$
 and at a different pressure $\beta_{p}=6.12$.
 Naturally, there are also 1-lobe solutions.
 For $n=4$ with $\Theta'(-1,4)$,
 Fig.5 shows a 3-lobe structure with $\beta_{p}=6.05$,
 which is rather symmetric as in Fig.2.
 However, the symmetric lobes
 of Fig.2 and Fig.5 are nonlinear lobes
 as they differ from the linear $\beta_{p}=0$
 corresponding lobes of Fig.1,
 where the (x,y) scales in the three figures
 have the same proportions.
 For $n=4$ and under the same boundary derivative $\Theta'(-1,4)$,
 Fig.6 shows a 2-lobe structure
 as pressure increases to $\beta_{p}=6.41$.
 In this figure, one lobe is overwhelmingly
 dominating over the other, as in Fig.3 and Fig.4,
 showing the nonlinear nature of the equation.
 For $n=4$ with $\Theta'(-1,3)$,
 Fig.7 shows an asymmetric 3-lobe structure
 with $\beta_{p}=12.3$.
 For $n=4$ and changing the boundary derivative to $\Theta'(-1,1)$,
 Fig.8 shows another asymmetric 3-lobe structure
 with $\beta_{p}=47.0$.
 Comparing the 3-lobe structures of $n=4$
 in Fig.5, Fig.7, and Fig.8,
 all with proportional (x,y) scales,
 the lobe amplitude decreases
 as the boundary derivatives decreases
 from $\Theta'(-1,4)$ to $\Theta'(-1,1)$.
 Multipole magnetic structures similar to ours
 have been explored in magneto-rotational supernovae
 \citep{bisnovatyi2008}.
 
We note that with $R(z)=1/z^{n}$
 the poloidal magnetic flux function
 is $\Psi(z,x)=R(z)\Theta(x)$.
 The poloidal magnetic field lines
 and the density weighed velocity stream lines
 with $\Psi(z,x)=C$
 are shown in Fig.9 for $n=3$, $\beta_{p}=6.12$,
 and in Fig.10 for $n=4$, $\beta_{p}=6.41$,
 corresponding to Fig.4 and Fig.6 respectively.
 Together with the toroidal component,
 they generate a surface of revolution
 about the polar axis.
 Taking the $2m(n)$ power of this poloidal flux function,
 and considering the complement contours,
 we could generate the mass density profiles of
 $\rho(\Psi)=\rho_{0}\bar\rho_{2}(\Psi)=\rho_{0}(C-|\Psi|^{2m(n)})$.
 The generalized total plasma pressure $\bar p_{*}$
 follows with an added $\bar p_{*1}=1/z$ decay with $z$.
 We remark that the nonlinear solution of $\Theta(x)$
 is driven by the generalized total pressure $\bar p_{*}$.
 According to this model,
 although $\Theta(x)$ is occasionally symmetric,
 this function is basically asymmetric
 and often with $\beta_{p}\gg 1$.
 Consequently, supernovae are asymmetric in nature
 with different degree of asymmetry.
 Observed from broadside,
 this asymmetry should be evident.
 As the line of sight moves towards the poles,
 the supernova becomes less asymmetrical
 and more spherical by projection effects.
 The proto-supernova cavity,
 filled with circulating plasmas and magnetic fields,
 generates a sequence of rotational equilibria
 as $\beta_{p}$ increases.
 When $\beta_{p}$ gets sufficiently high
 to crack the stellar hydrogen envelope,
 an often asymmetric supernova is erupted
 resulting in the recoil of the neutron star.
 Since $\beta\gg 1$, plasma pressure
 is the primary trigger of supernova
 rather than magnetic field,
 which is consistent to some simulations
 \citep{komissarov2007}.

\newpage
\section{Collimated Polar GRB Jets}

We can also arrange the Grad-Shafranov equation,
 Eq.~\ref{eqno22}, as

\begin{eqnarray}
\nonumber
r^{2}{1\over R}{d^2R\over dr^2}
 \pm m\beta_{p}z^{4}\sin^{2}\theta
 \bar p_{*1}(z)R^{2m-2}|\Theta|^{2m-2}\,
 &=&\,-{1\over \Theta}\sin\theta{d\over d\theta}
 \left({1\over\sin\theta}{d\Theta\over d\theta}\right)\,
\\
\label{eqno28}
 &=&\,n(n+1)\,\,\,.
\end{eqnarray}

\noindent The $\Theta$ equation and its solutions are 

\begin{mathletters}
\begin{eqnarray}
\label{eqno29a}
(1-x^2){d^2\Theta(x)\over dx^2}
 +n(n+1)\Theta(x)\,
 =\,0\,\,\,,
\\
\label{eqno29b}
\Theta(x)\,=\,(1-x^2){d P_{n}(x)\over dx}\,
 =\,(1-x^2)\,\,\,,
\end{eqnarray}
\end{mathletters}

\noindent where $P_{n}(x)$ is the Legendre polynomials
 and the last equality is for $n=1$.
 As for the $R$ equation, it reads

\begin{eqnarray}
\nonumber
r^{2}{d^2R\over dr^2}-n(n+1)R\,
 &=&\,\mp m\beta_{p}z^{4}\bar p_{*1}(z)R^{2m-1}
 \sin^{2}\theta|\Theta|^{2m-2}\,
\\
\label{eqno30}
 &=&\,\mp m\beta_{p}z^{3}R^{2m-1}
 \sin^{2}\theta|\Theta|^{2m-2}\,\,\,,
\end{eqnarray}

\noindent where we have made use of Eq.~\ref{eqno18}
 to get the last equality.
 To make this equation separable,
 we consider $2m-2=-1$ and then take $n=1$
 in Eq.~\ref{eqno29b} to get

\begin{eqnarray}
\label{eqno31}
r^{2}{d^2R\over dr^2}-n(n+1)R\,
 =\,\mp m\beta_{p}z^{3}R^{2m-1}
 {\sin^{2}\theta\over\Theta}\,
 =\,\mp m\beta_{p}z^{3}\,\,\,.
\end{eqnarray}

\noindent With $n=1$ and $2m=1$, the solution of $R(z)$
 is given by the homogeneous solution in bracket
 plus the particular solution

\begin{eqnarray}
\label{eqno32}
R(z)\,&
 =&\,\left(az^{n+1}+b{1\over z^{n}}\right)+cz^{3}\,\,\,,
\\
\nonumber
c\,&=&\,\mp{1\over 4} m\beta_{p}\,\,\,.
\end{eqnarray}

To solve for equilibrium $R(z)$,
 we again take the lower sign in Eq.~\ref{eqno31}.
 Taking $a=+2$, $b=+2$, $c=+\beta_{p}/8$, and $\beta_{p}=5$,
 Fig.11 shows the function $R(z)$.
 Together with $\Theta(x)=(1-x^{2})$,
 the poloidal magnetic field lines
 and the poloidal density weighed velocity stream lines,
 $\Psi(z,x)=R(z)\Theta(x)=C$,
 run on the contour line as shown in Fig.12
 with $C=2,3,4,5,6,7$.
 Together with the toroidal component,
 they generate a surface of revolution
 about the polar axis.
 The magnetic field lines of this jet in the cavity
 can be closed via the stellar envelope.
 For large distances,
 the surface is dominated by the $z^{3}$ term
 with $\Psi(z,x)=cz^{3}(1-x^{2})=zz^{2}_{\perp}=C$,
 where $z_{\perp}$ is the perpendicular distance
 from the polar axis.
 Let us approximate the contour as

\begin{eqnarray}
\label{eqno33}
z^{2}_{\perp}\,=\,{C\over z_{\parallel}}\,\,\,.
\end{eqnarray}

\noindent This shows that as $z$ increases,
 $z_{\perp}$ would decrease,
 giving a cusp funnel structure,
 or a tornado-like line vortice structure,
 of a polar jet.
 In Fig.12, only one jet is shown for more clarity,
 and the opposing jet is likewise.
 Furthermore, only the initial part
 of the funnel contours are shown.
 For $y(\theta=0)>15$, the analytic funnel solution
 of Eq.~\ref{eqno33} should be superimposed on it.
 We also note that the inner contours 2,3 and 4
 are connected to the outer magnetosphere,
 while the outer contours 5,6, and 7
 would have to come from external sources,
 like an accretion disk.
 If these external sources are unable to provide
 such corresponding field strengths of the contours,
 these outer jet contours should be ignored.
 For small distances,
 the poloidal flux surface is dominated
 by the $b/z$ term with $\Psi(z,x)=b(1-x^{2})/z=C$,
 such that $z$ is proportional to $\sin^{2}\theta$.
 Considering $\sin^{2}\theta$
 as the dipole field of the magnetar,
 the polar funnel structure connects up
 with the magnetosphere of the central compact star,
 as is shown in a close-up look in Fig.13.
 We name this as the magnetospheric jet.
 This connection should take place
 outside the light cylinder of the magnetosphere.

Furthermore, for a given $z_{\parallel}$,
 Eq.~\ref{eqno33} shows that $z^{2}_{\perp}$
 is proportional to the contour value $C$.
 As a result, the gradient of $\Psi$ increases outward.
 This means stronger field lines
 and faster stream lines are on the outside.
 Because of the overall pressure balance,
 higher pressure plasmas are on the inside close to the axis.
 Plasma confinement is, therefore,
 accomplished by the strong magnetic fields outside
 surrounding the plasmas.
 The magnetic confinement is also enhanced
 by the inward magnetic curvature
 of the cusp funnel geometry
 which puts a magnetic tension force on the plasmas.
 As for the mass density,
 which satisfies $\vec v\cdot\nabla\rho(\Psi)=0$,
 the density gradient $\nabla\rho(\Psi)=(d\rho/d\Psi)\nabla\Psi$
 is perpendicular to the $\Psi$ contour surface.
 The mass density profiles
 are given by the complement contours
 of the poloidal flux function, according to 
 $\rho(\Psi)=\rho_{0}(C-|\Psi|^{2m})=\rho_{0}(C-|\Psi|)$
 with $2m=1$,
 and the generalized total plasma pressure $\bar p_{*}$
 follows accordingly.
 Although the cusp funnel volume is unbounded,
 the jet will be filled up to the point
 where the energy flux
 of the star's magnetosphere can supply over time.
 Upon plowing through the stellar envelope
 along the polar axis,
 this cavity jet could then erupt
 into a gamma-ray burst jet.

According to our findings,
 the outer magnetosphere
 is torn open by the plasma ram pressure,
 and is flipped polewards
 by the poloidal plasma flows
 to form a cusp funnel.
 Consequently, the magnetic field lines of this funnel
 wrap around a high pressure plasma column around the axis.
 Usually, when only toroidal rotational plasma velocity
 is taken into account,
 the outer magnetosphere opens up
 into monopole-like radial fields.
 In our case, the poloidal plasma velocity
 brings the outer magnetosphere to the polar direction.
 This offers a MHD description
 of the magnetic towering mechanism
 \citep{lynden2003}.
 Standard jet formation schemes rely on
 the angular momentum of a binary system
 or an accretion disk-compact star system
 \citep{prendergast2005, gourgouliatos2010}.
 A single star discounting the external stellar envelope
 can hardly provide the needed angular momentum to form jets.
 The present MHD model can provide asymmetric supernova
 and polar collimating gamma-ray burst jet
 configurations in the cavity,
 therefore, favoring the gamma-ray burst and supernova association
 \citep{galama1999, campana2006, pian2006, soderberg2006, mazzali2006}.
 Our axial jet has a specific "beam pattern".
 Such feature is compatible with models
 that consider gamma-ray burst and afterglow properties
 as a result of viewing angle
 on the beam pattern of the jet
 \citep{lipunov2001, rossi2002, salmonson2002}.

\newpage
\section{AGN Plasma Torus and Conclusions}

Active galactic nuclei (AGN) is a term
 used to designate astrophysical objects
 of Seyferts, quasars, radio galaxies, and blazers (BL-Lacs),
 that emit mighty electromagnetic radiations
 at different bands from radio, optical, to X-ray, gamma-ray.
 From the spectroscopic characteristics of these objects,
 it is deduced that they could be the results of a single
 unified AGN structure viewed at different angles.
 This unified AGN model \citep{antonucci1993}
 would consist of a central core with a black hole candidate,
 an acretion disk, that defines the equatorial plane,
 and a dust torus further out on this plane.
 On the rotational (polar) axis of the central core,
 there would be two opposing jets that emit strongly beamed
 and polarized radio waves.
 Viewed from face-on (down the polar axis)
 would give blazers,
 and from edge-on (on the equatorial plane)
 would give type 2 Seyferts, quasars,
 and radio galaxies with narrow emission lines.
 While viewed at an intermediate angle
 would give type 1 Seyferts, quasars,
 and radio galaxies with broad emission lines.
 The material in the acretion disk gravitates inward,
 by transporting its angular momentum outward through dissipations,
 and feeds the central core to generate the jets.
 The existence of a dust torus in AGN has been inferred
 from high resolution instruments and images.
 
To account for this unified AGN structure,
 we consider another solution of Eq.~\ref{eqno32}
 with a negative coefficient $a$.
 Writing the negative sign explicitly,
 we have with $n=1$

\begin{eqnarray}
\label{eqno34}
R(z)\,
 =\,-az^{n+1}+\left(b{1\over z^{n}}+cz^{3}\right)\,\,\,.
\end{eqnarray}

\noindent The two terms in bracket
 generates a minimum in the positive $R(z)$ domain.
 The first term shifts the minimum
 down to the negative domain
 bounded by $z_{1}$ and $z_{2}$
 where $R(z_{1})=R(z_{2})=0$.
 With the negative sign explicit,
 we take again $a=+2$, $b=+2$, $c=+\beta_{p}/8$,
 and $\beta_{p}=5$,
 the function $R(z)$ is shown in Fig.14 with three regions,
 separated by $z_{1}=1.17$ and $z_{2}=3.09$.
 In the absence of the envelope,
 the poloidal flux function $\Psi(z,x)=R(z)\Theta(x)$
 with $\Theta(x)=(1-x^{2})$ is shown in Fig.15,
 which consists of a dipole-like magnetosphere,
 a plasma torus, and a polar jet
 separated by two spherical separatrix at $z_{1}$ and $z_{2}$.
 This polar jet can be formed
 if there is horizontal accretion to feed it.
 We name this as the accretion jet
 in contrast to the magnetospheric jet.
 A close-up look at the AGN magnetosphere
 is shown in Fig.16,
 which gives an intrinsic AGN magnetic moment
 \citep{schild2006}.
 Naturally, upon this basic AGN structure,
 we can superimpose an AGN magnetospheic jet
 as discussed in the preceeding section
 to generate quasars and blazers.
 
To conclude, we have used a set of MHD equations
 with divergence-free axisymmetric
 poloidal and toroidal non-field aligned plasma flows
 to study the dynamics of the cavity
 between the central core and the external envelope
 in the final phase of a collapsing star.
 A sequence of steady state rotational MHD equilibria
 in response to the increasing plasma pressure
 is solved to represent the plasma evolution in the cavity.
 The spatial configuration is described
 by the rotational Grad-Shafranov equation
 where the ratio of the generalized total plasma pressure
 to the poloidal magnetic pressure,
 $\beta_{p}=2\mu p_{0}/(a^{2}A_{0})^{2}$, is the cavity parameter.
 By assigning two source functions,
 the rotational Grad-Shafranov equation
 can be solved for asymmetric supernova,
 polar collimated cusp funnel gamma-ray burst jet,
 and active galactic nucleus plasma torus.
 It is important to remark
 that both the asymmetric supernova lobes
 and the cusp gamma-ray burst polar jets
 are connected directly to the magnetosphere
 of the central compact star,
 not to an accretion disk.
 This structure identifies the magnetosphere
 as the central engine
 of supernova and gamma-ray burst events,
 by providing plasma, magnetic energy,
 and rotational energy.
 Since $\beta_{p}$ gets much larger than unity,
 plasma pressure is likely to be the primary agent
 in cracking the stellar envelope,
 instead of the magnetic field.
 
\acknowledgments

\appendix

\clearpage
\begin{figure}
\plotone{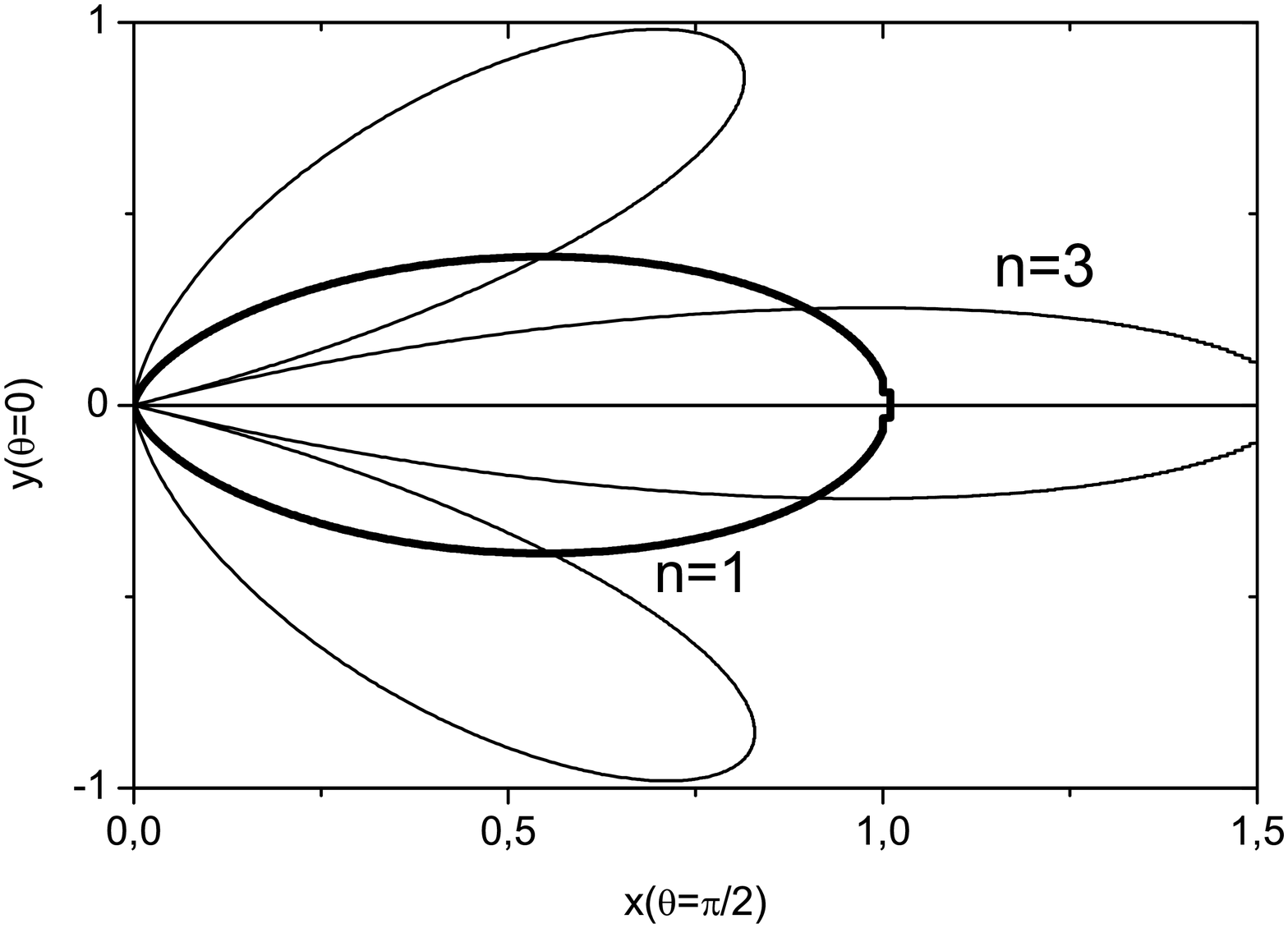}
\caption{The lobe structures of $\Theta(x)$
 for $n=1,2,3$ with $\beta_{p}=0$
 are plotted on the $(r-\theta)$ plane.}
\label{fig.1}
\end{figure}

\clearpage
\begin{figure}
\plotone{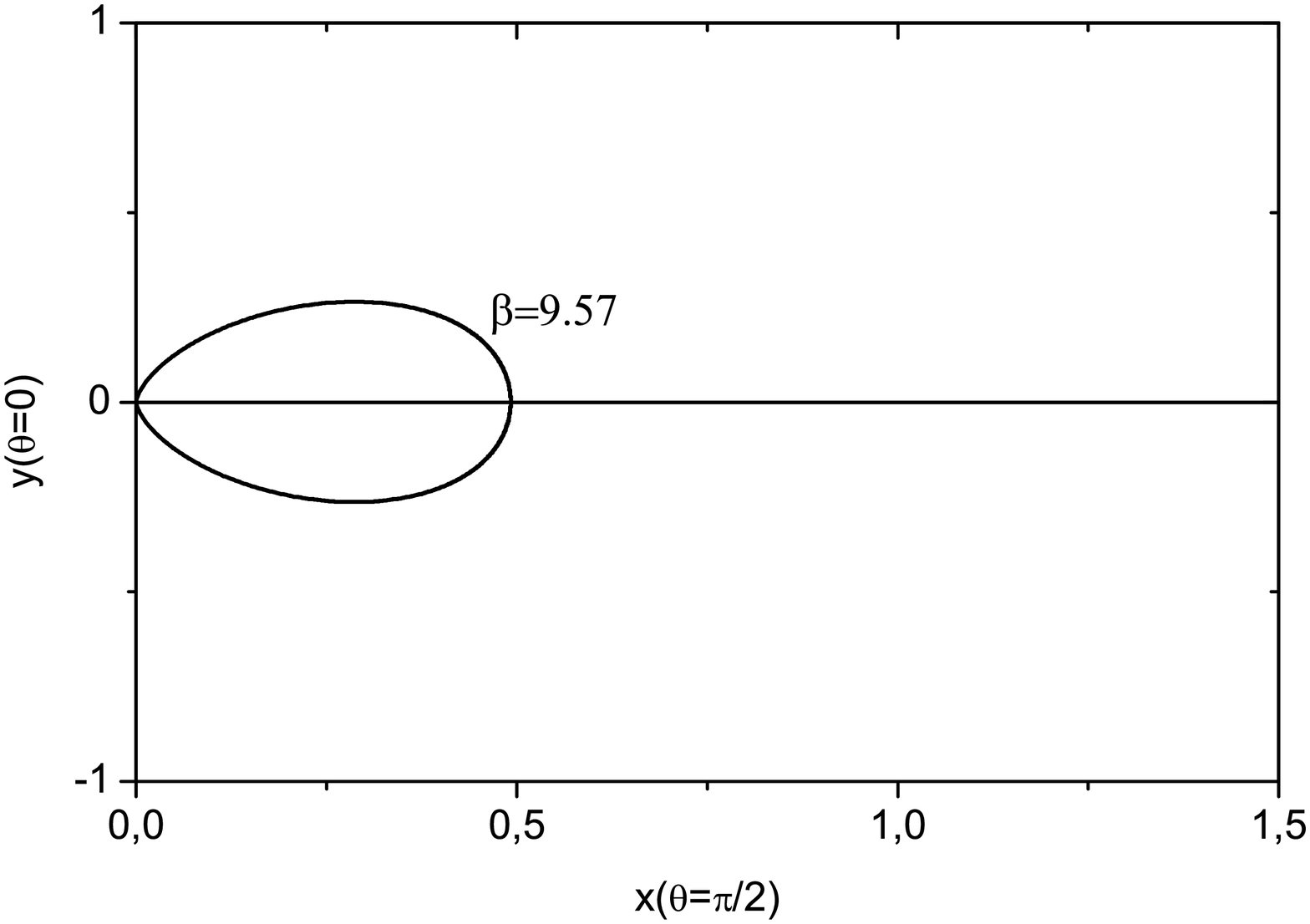}
\caption{The 1-lobe structure of $\Theta(x)$
 for $n=2$ with $\beta_{p}=9.57$ and $\Theta'(-1,1)$
 is plotted on the $(r-\theta)$ plane.}
\label{fig.2}
\end{figure}

\clearpage
\begin{figure}
\plotone{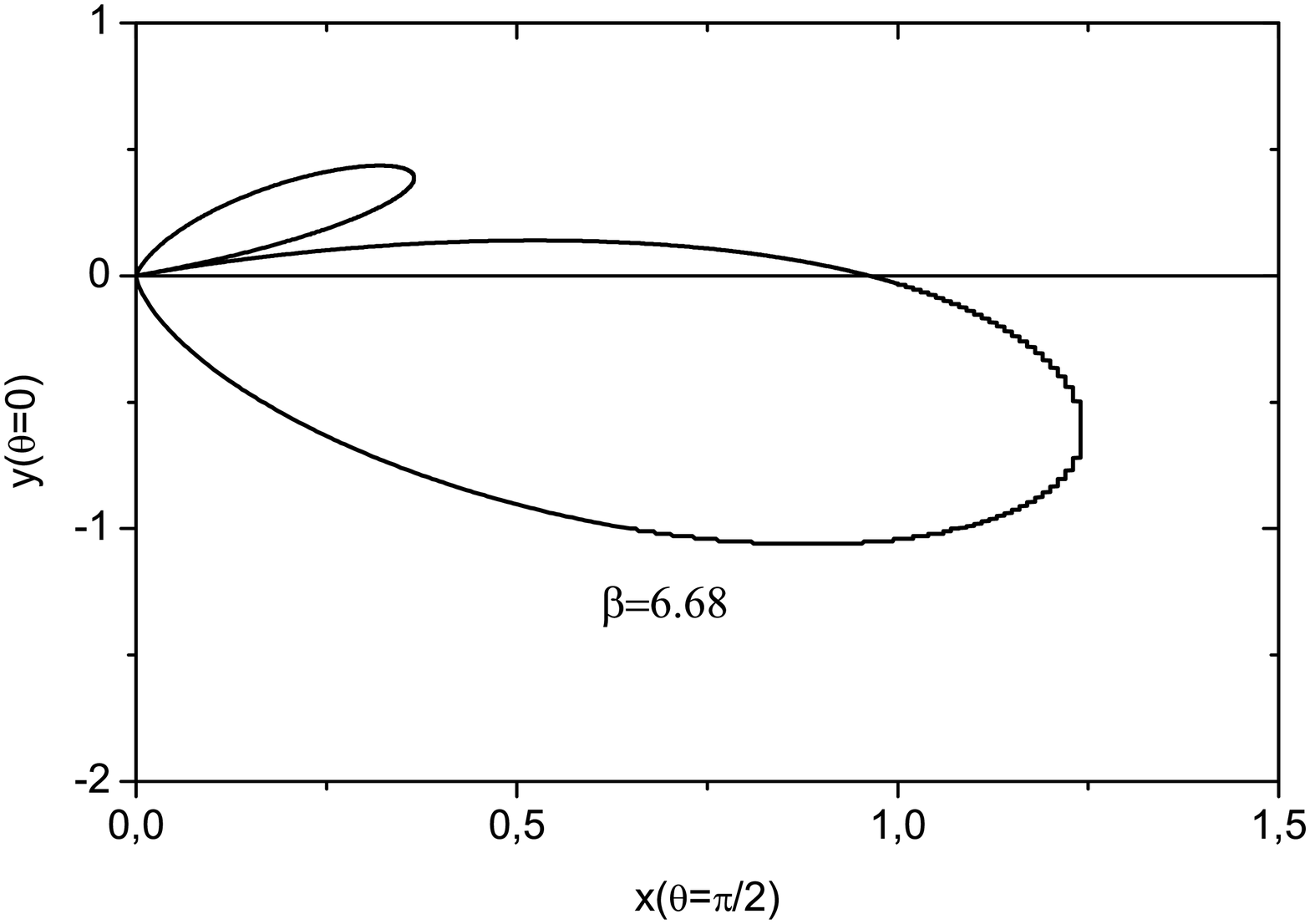}
\caption{The 2-lobe structure of $\Theta(x)$
 for $n=3$ with $\beta_{p}=6.69$ and $\Theta'(-1,3)$
 is plotted on the $(r-\theta)$ plane.}
\label{fig.3}
\end{figure}

\clearpage
\begin{figure}
\plotone{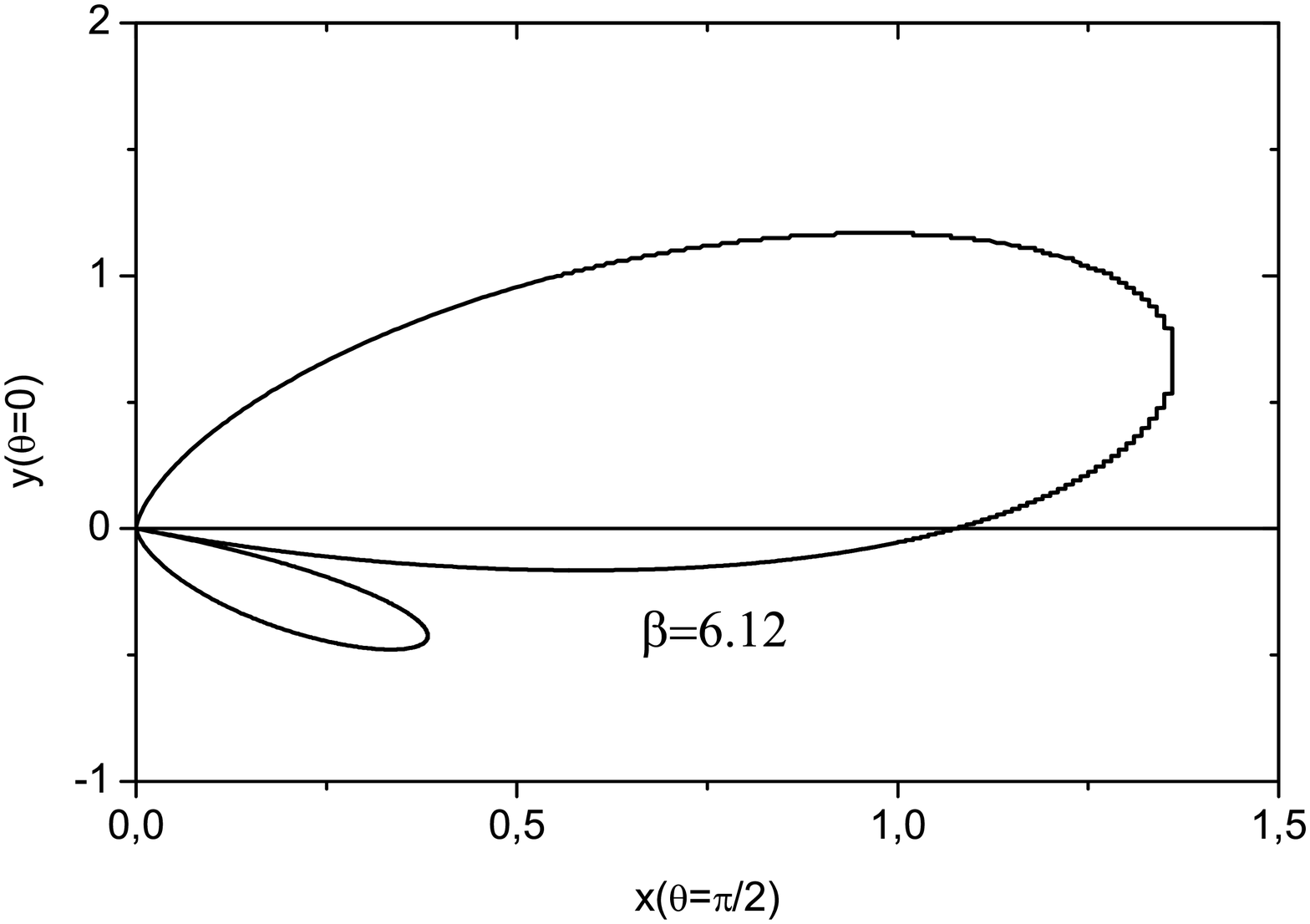}
\caption{The 2-lobe structure of $\Theta(x)$
 for $n=3$ with $\beta_{p}=6.12$ and $\Theta'(-1,2)$
 is plotted on the $(r-\theta)$ plane.}
\label{fig.4}
\end{figure}

\clearpage
\begin{figure}
\plotone{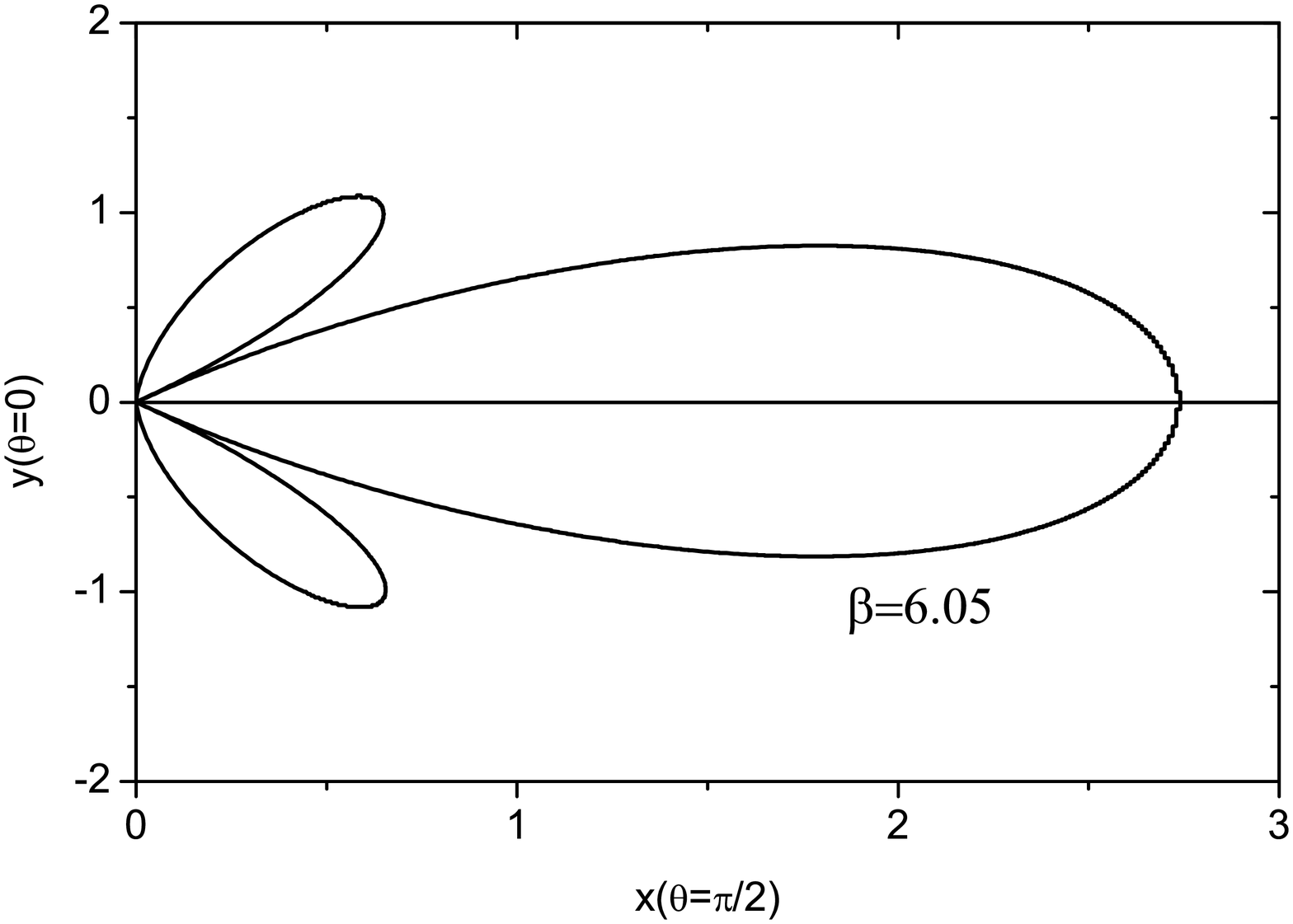}
\caption{The 3-lobe structure of $\Theta(x)$
 for $n=4$ with $\beta_{p}=6.05$ and $\Theta'(-1,4)$
 is plotted on the $(r-\theta)$ plane.}
\label{fig.5}
\end{figure}

\clearpage
\begin{figure}
\plotone{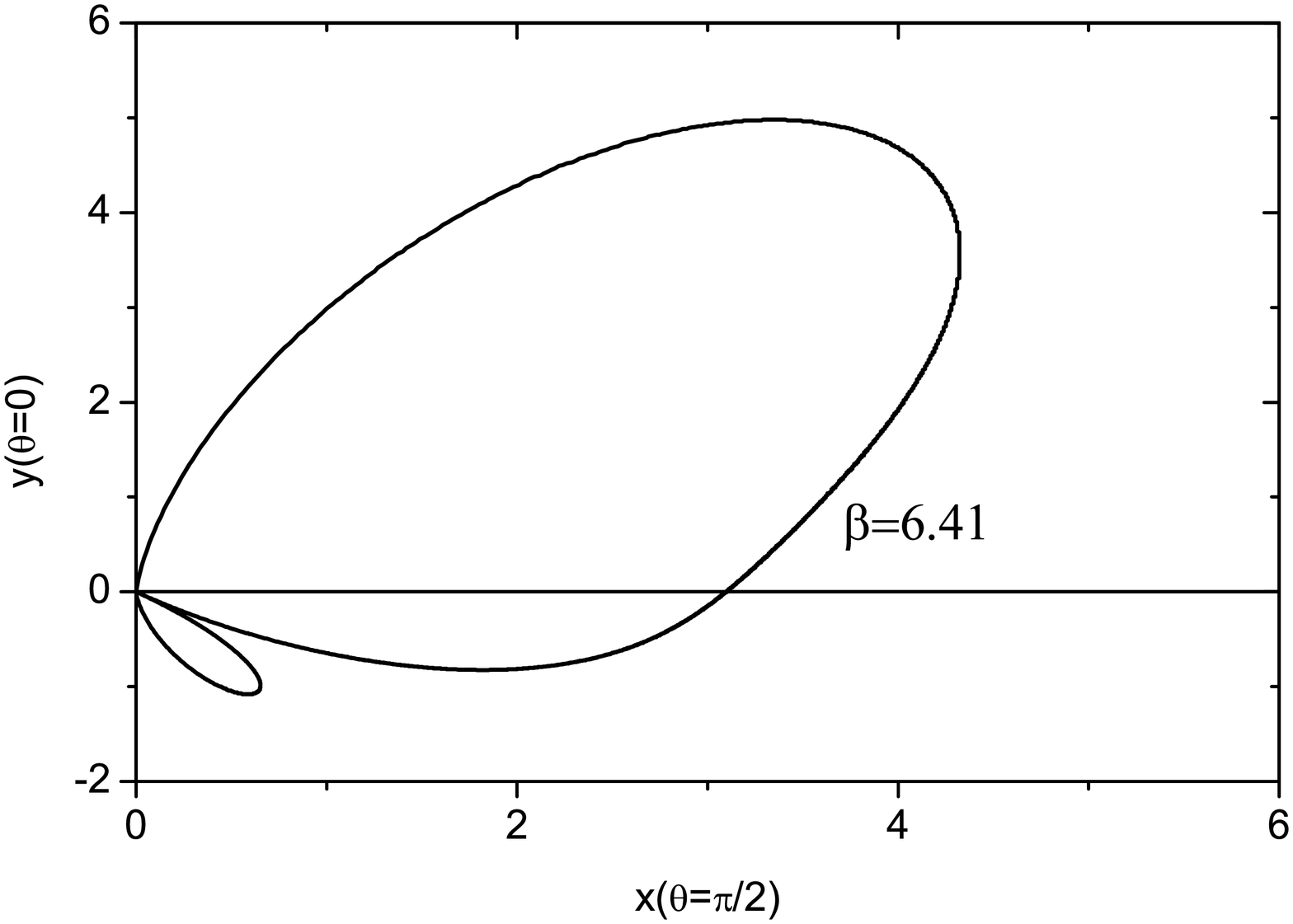}
\caption{The 2-lobe structure of $\Theta(x)$
 for $n=4$ with $\beta_{p}=6.41$ and $\Theta'(-1,4)$
 is plotted on the $(r-\theta)$ plane.}
\label{fig.6}
\end{figure}

\clearpage
\begin{figure}
\plotone{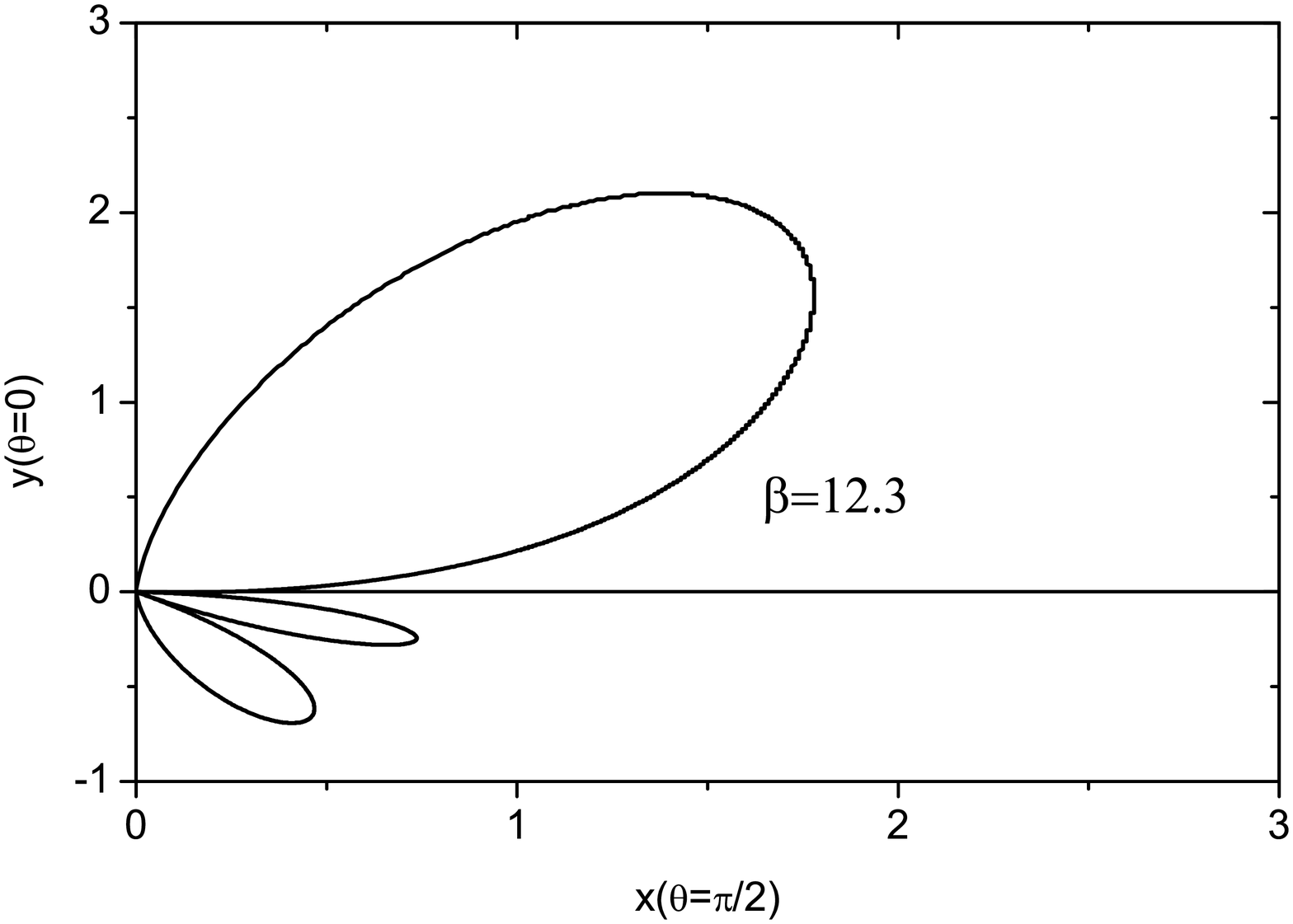}
\caption{The 3-lobe structure of $\Theta(x)$
 for $n=4$ with $\beta_{p}=12.3$ and $\Theta'(-1,3)$
 is plotted on the $(r-\theta)$ plane.}
\label{fig.7}
\end{figure}

\clearpage
\begin{figure}
\plotone{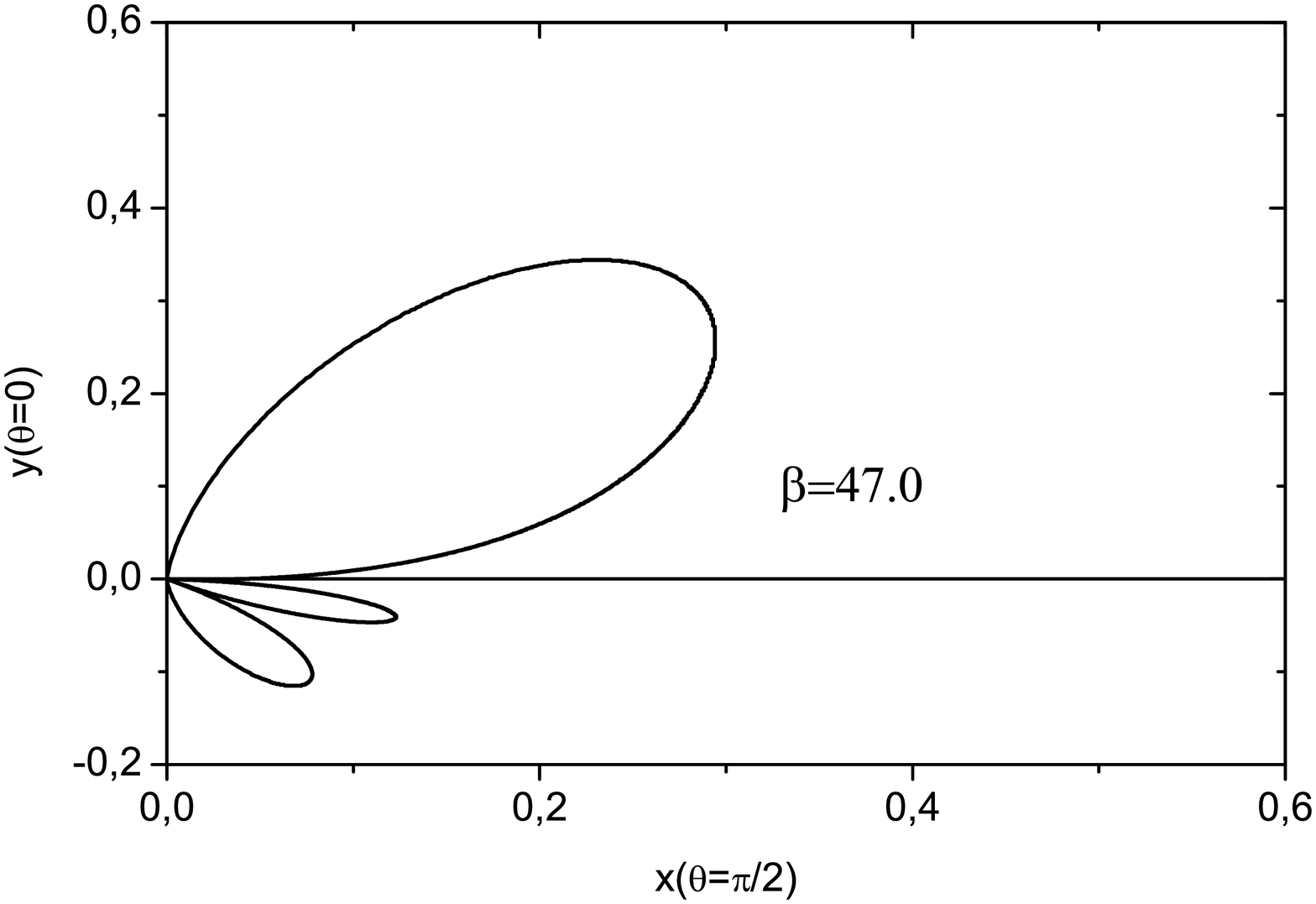}
\caption{The 3-lobe structure of $\Theta(x)$
 for $n=4$ with $\beta_{p}=47.0$ and $\Theta'(-1,1)$
 is plotted on the $(r-\theta)$ plane.}
\label{fig.8}
\end{figure}

\clearpage
\begin{figure}
\plotone{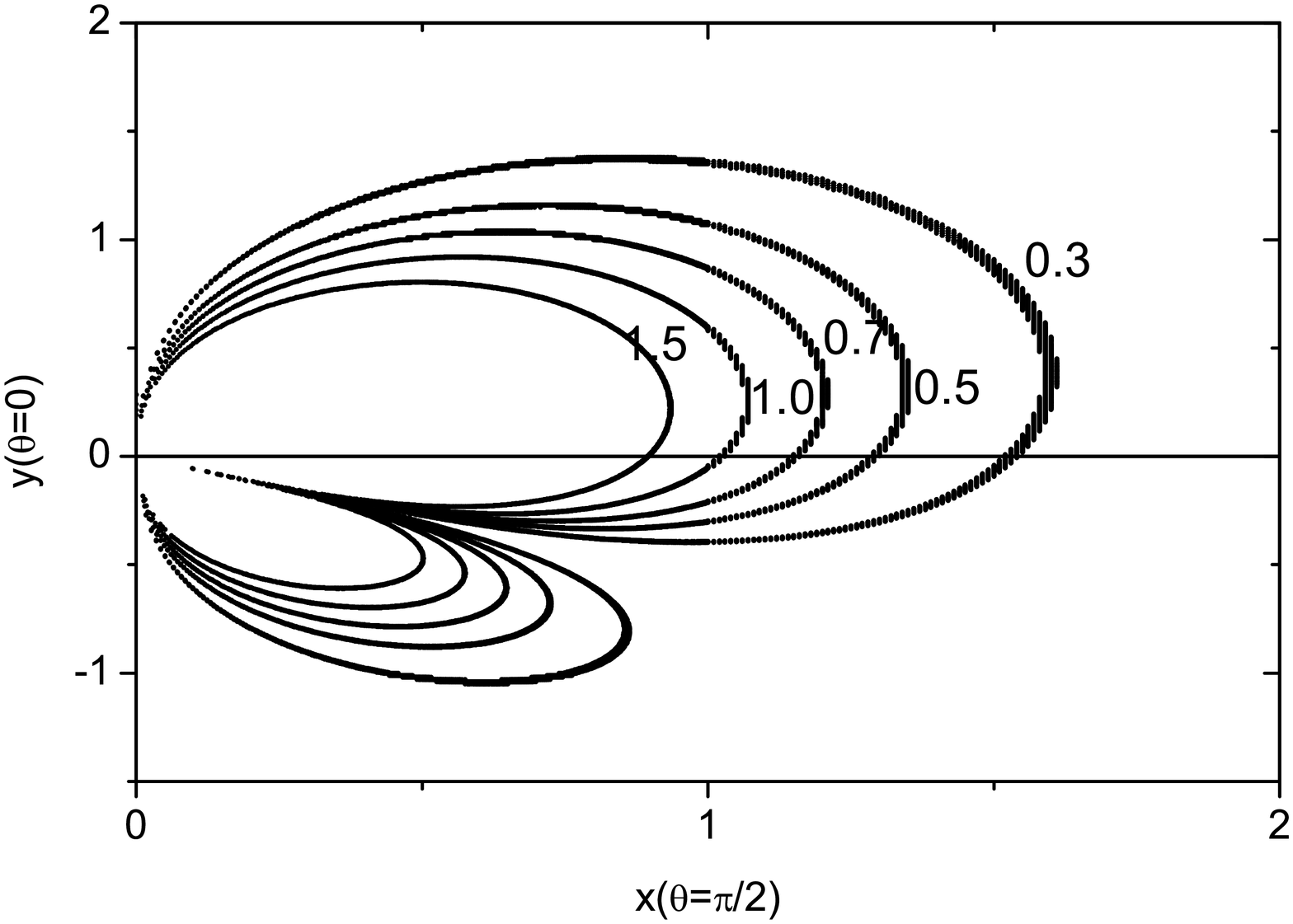}
\caption{The poloidal flux function $\Psi(z,x)$ contours
 for $n=3$ with $\beta_{p}=6.12$
 are plotted on the $(r-\theta)$ plane
 with contour values of $C=1.5,1.0,0.7,0.5,0.3$
 to show the asymmetric supernova poloidal magnetic field lines
 and density weighed plasma velocity stream lines.}
\label{fig.9}
\end{figure}

\clearpage
\begin{figure}
\plotone{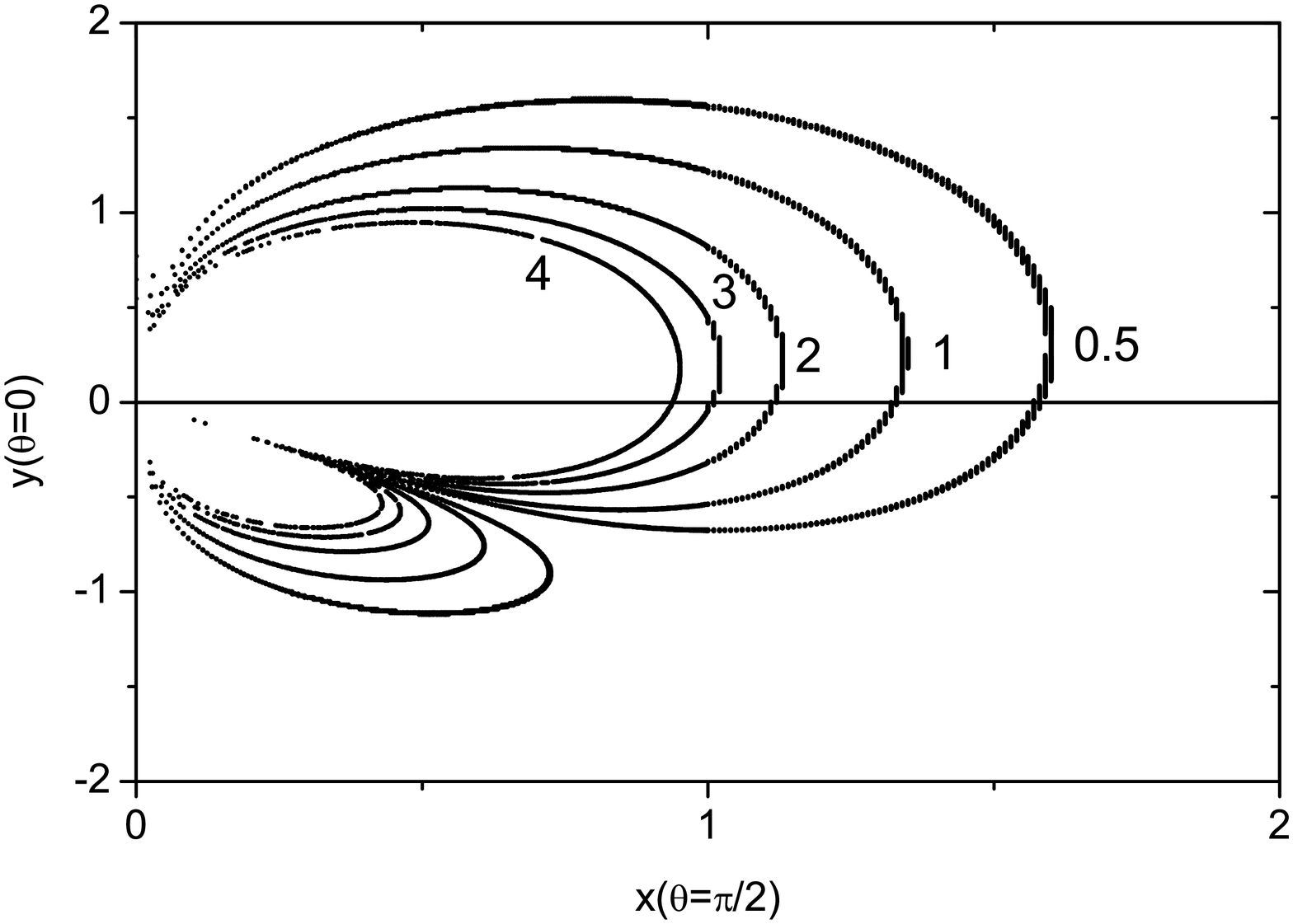}
\caption{The poloidal flux function $\Psi(z,x)$ contours
 for $n=4$ with $\beta_{p}=6.41$
 are plotted on the $(r-\theta)$ plane
 with contour values of $C=4,3,2,1,0.5$
 to show the asymmetric supernova poloidal magnetic field lines
 and density weighed plasma velocity stream lines.}
\label{fig.10}
\end{figure}

\clearpage
\begin{figure}
\plotone{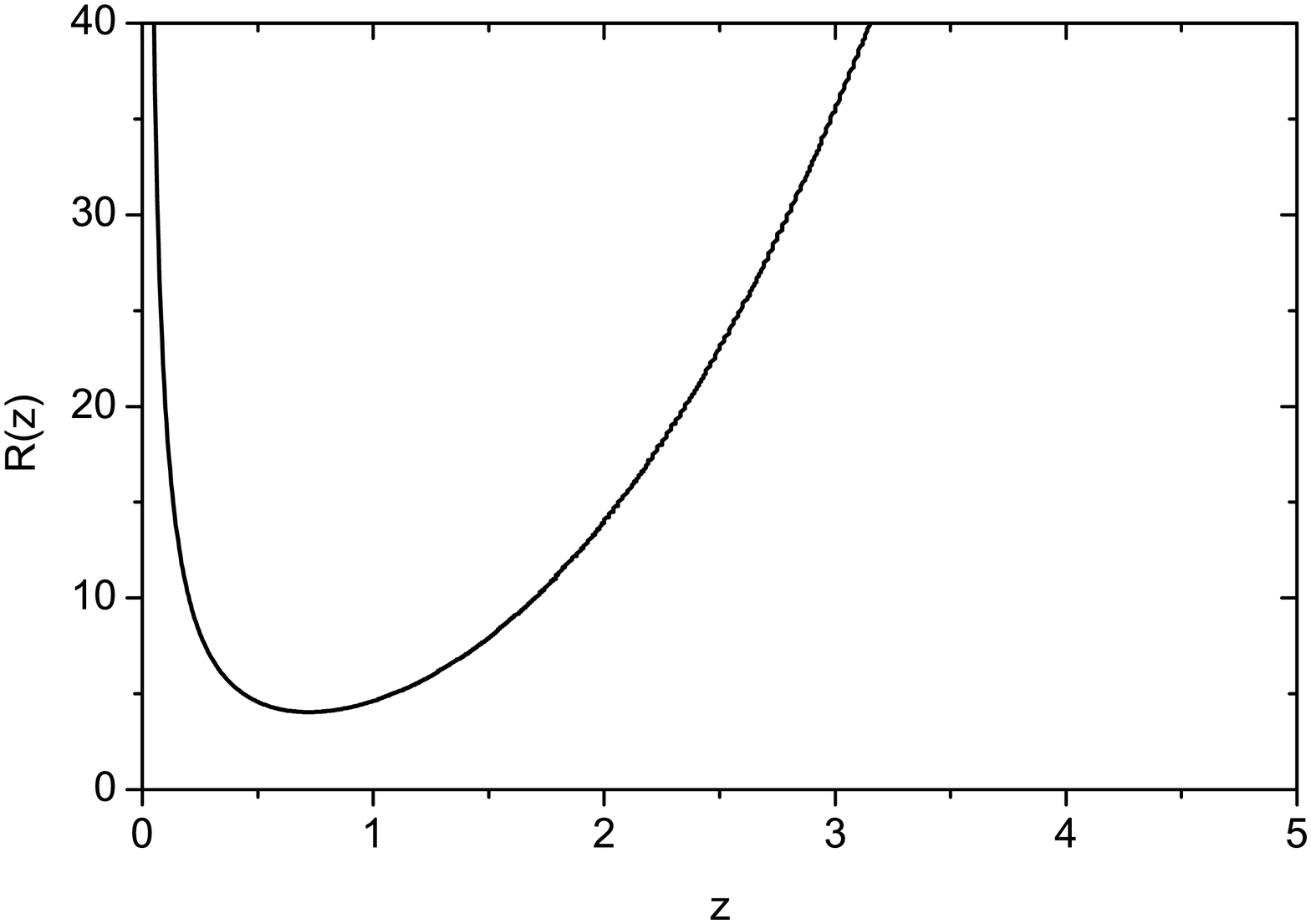}
\caption{The profile of $R(z)$
 is plotted with $a=+2$, $b=+2$, $c=+\beta_{p}/8$,
 and $\beta_{p}=5$.}
\label{fig.11}
\end{figure}

\clearpage
\begin{figure}
\plotone{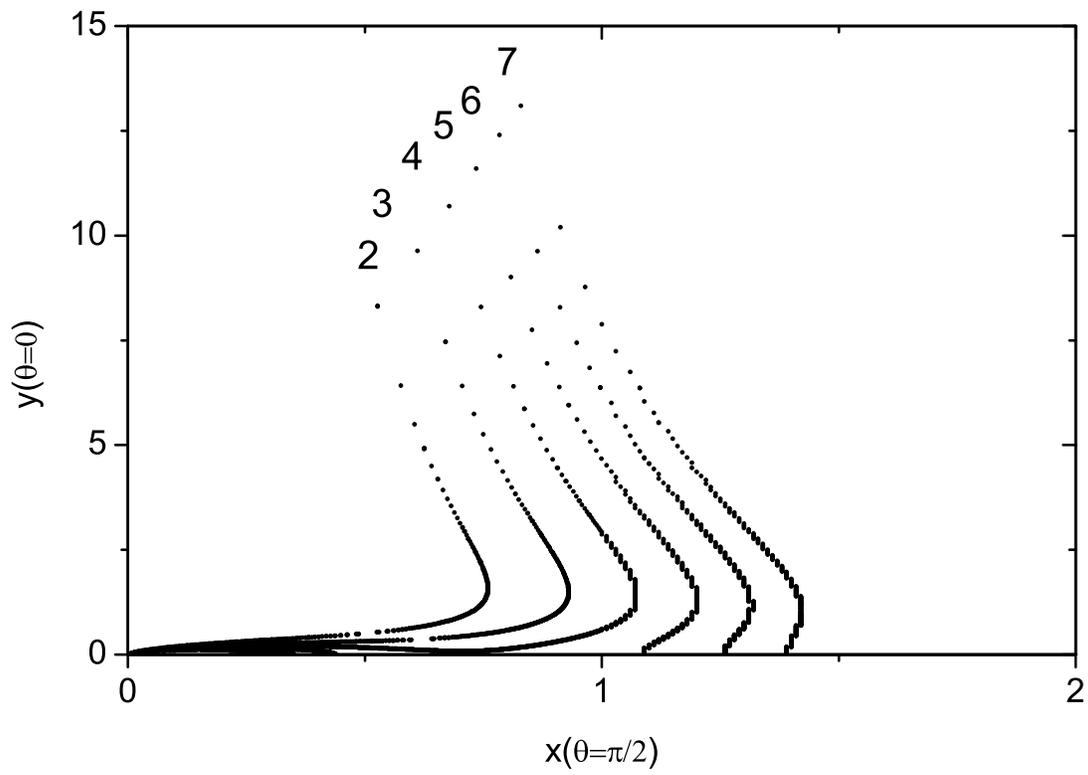}
\caption{The poloidal flux function $\Psi(z,x)$ contours
 are plotted on the $(r-\theta)$ plane
 with contour values of $C=7,6,5,4,3,2$
 to show the colimating GRB cusp funnel along the polar axis.}
\label{fig.12}
\end{figure}
 
\clearpage
\begin{figure}
\plotone{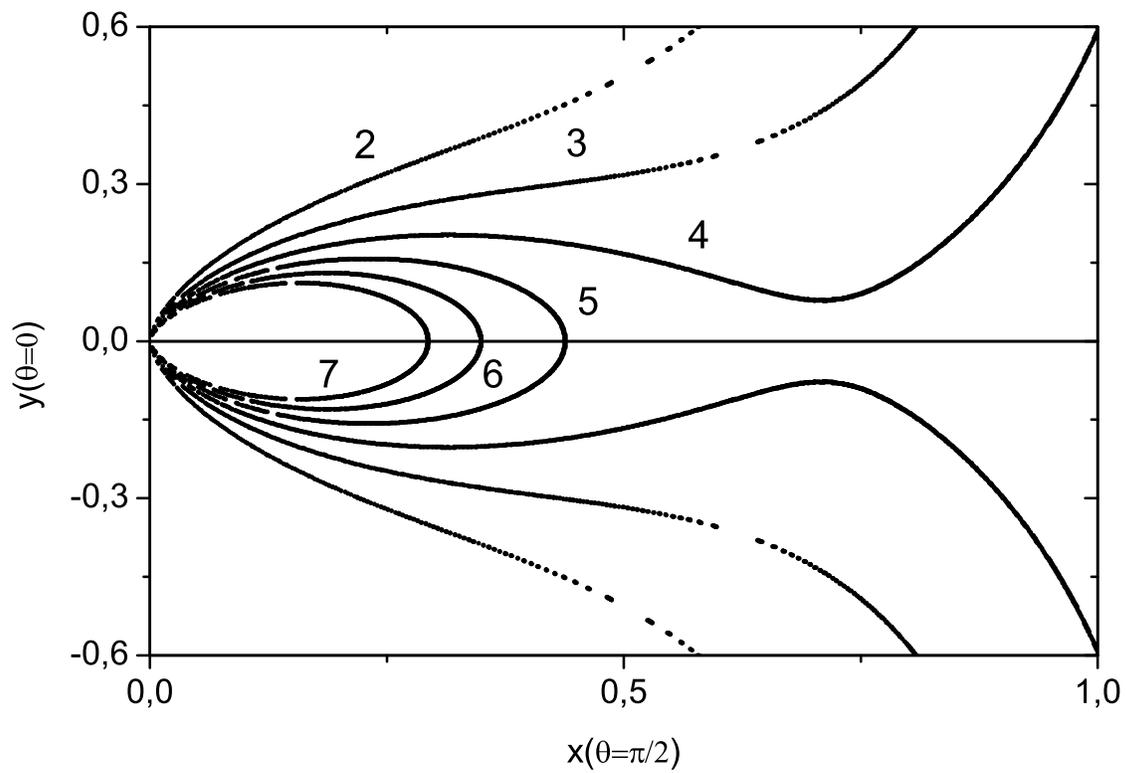}
\caption{The poloidal flux function $\Psi(z,x)$ contours
 are plotted on the $(r-\theta)$ plane
 with contour values of $C=7,6,5,4,3,2$
 to show the openning of the magnetosphere to the polar cusp.}
\label{fig.13}
\end{figure}

\clearpage
\begin{figure}
\plotone{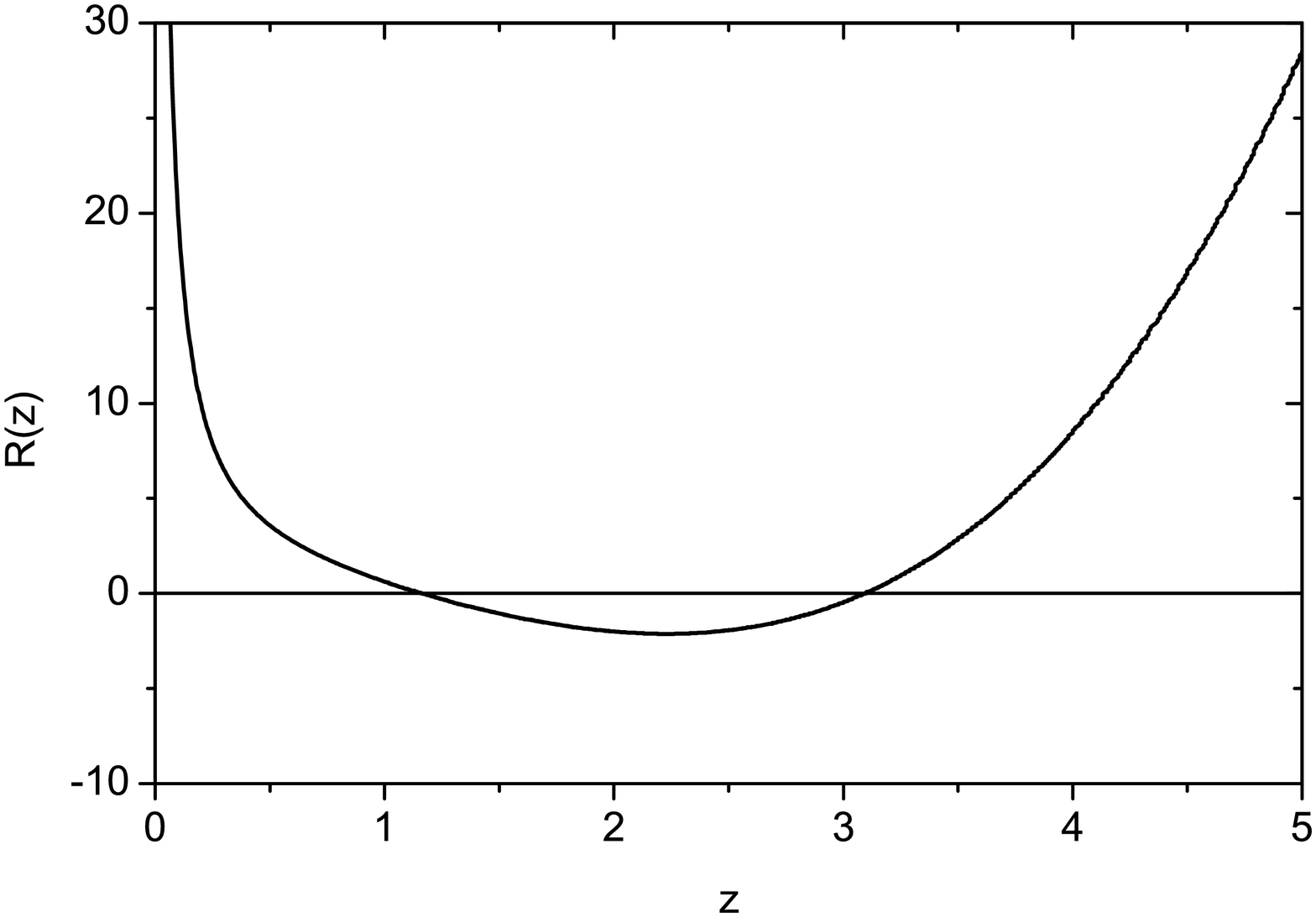}
\caption{The profile of $R(z)$ with a negative domain
 is plotted with $a=+2$, $b=+2$, $c=+\beta_{p}/8$,
 and $\beta_{p}=5$.}
\label{fig.14}
\end{figure}

\clearpage
\begin{figure}
\plotone{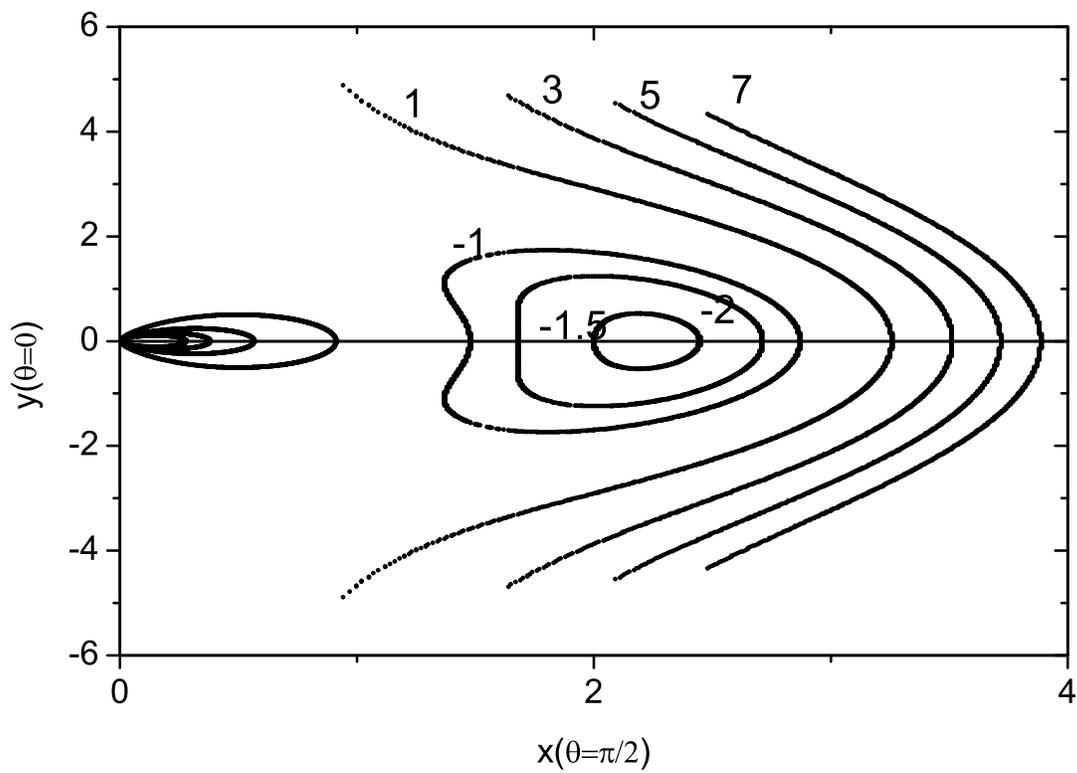}
\caption{The poloidal flux function $\Psi(z,x)$ contours
 are plotted on the $(r-\theta)$ plane
 to show the AGN magnetosphere,
 plasma torus with $C=-1,-1.5,-2$,
 and polar jet with $C=7,5,3,1$,
 separated by two spherical separatrix
 at $z_{1}=1.17$ and $z_{1}=3.09$.}
\label{fig.15}
\end{figure}

\clearpage
\begin{figure}
\plotone{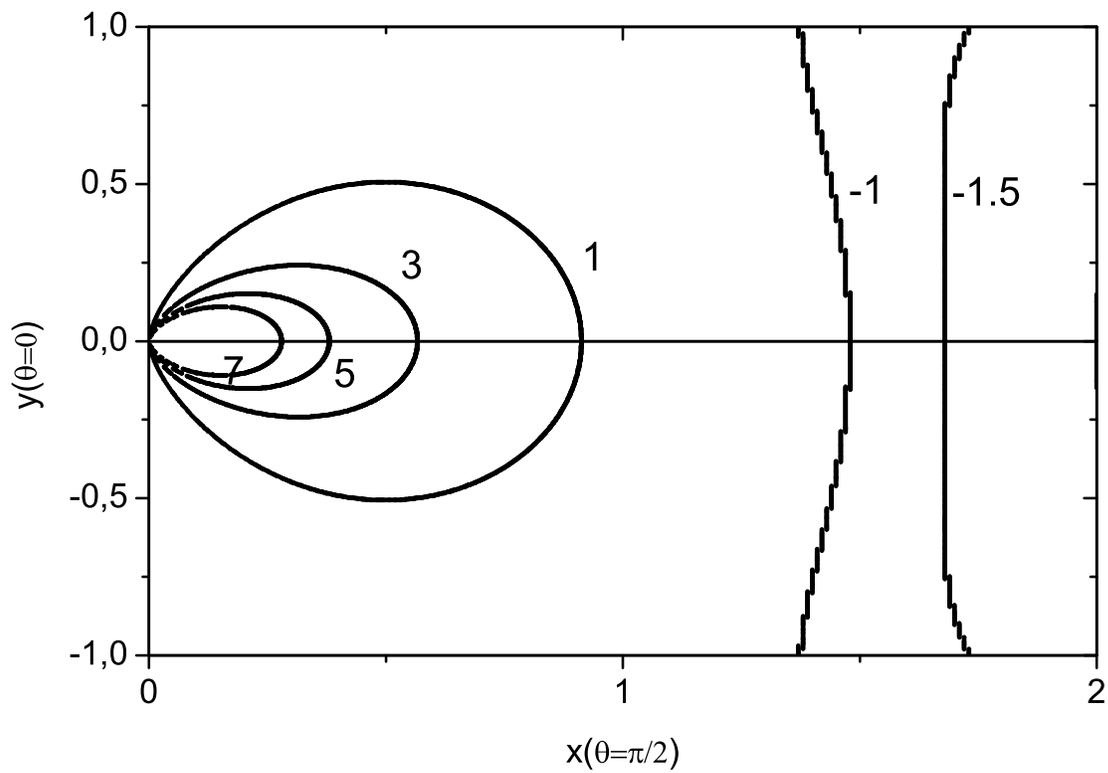}
\caption{The poloidal flux function $\Psi(z,x)$ contours
 are plotted on the $(r-\theta)$ plane
 to show the AGN magnetosphere with $C=7,5,3,1$.}
\label{fig.16}
\end{figure}

\end{document}